\documentstyle[11pt]{article} 
\def\baselinestretch{1.3}
\newcommand{\ba}{\begin{array}}
\newcommand{\ea}{\end{array}}
\newcommand{\bd}{\begin{displaymath}}
\newcommand{\ed}{\end{displaymath}}
\newcommand{\be}{\begin{equation}}
\newcommand{\ee}{\end{equation}}
\newcommand{\bea}{\begin{eqnarray}}
\newcommand{\eea}{\end{eqnarray}}

\def\bra{\langle}
\def\ket{\rangle}

\def\a{\alpha}
\def\as {\alpha_s}
\def\b{\beta}
\def\g{\gamma}
\def\d{\delta}
\def\e{\epsilon}
\def\ve{\varepsilon}
\def\l{\lambda}
\def\m{\mu}
\def\n{\nu}
\def\G{\Gamma}
\def\D{\Delta}
\def\L{\Lambda}
\def\s{\sigma}
\def\p{\pi}

\def\mzs {M_Z^2}
\def\mws {M_W^2}
\def\q2 {q^2}
\def\sz {\sin^2\theta_W}
\def\cz {\cos^2\theta_W}

\def\r {\rightarrow}
\def\t {\times }
\def\slash {\!\!\!\!\!\!/}
\def\photino {\tilde\gamma}
\def\neu {\chi_1^0}
\def\rslep {\tilde{e_R}}
\def\lslep {\tilde{e_L}}
\def\mrslep {m_{\rslep}}
\def\mlslep {m_{\lslep}}
\def\mneu {m_{\neu}}

\begin{document}
\begin{flushright}
{\large MRI-PHY/P970923\\ September 1997\\ hep-ph/9709431}
\end{flushright}

\begin{center}
{\Large\bf
Gauge-Mediated Supersymmetry Breaking Signals in an 
Electron-Photon Collider}\\[20mm]
{Ambar Ghosal}\footnote{Electronic address: ambar@mri.ernet.in}, 
{Anirban Kundu}\footnote{Electronic address: akundu@mri.ernet.in}
and {Biswarup Mukhopadhyaya}\footnote{Electronic address: 
biswarup@mri.ernet.in}\\[10mm]
{\em Mehta Research Institute,\\
Chhatnag Road, Jhusi, Allahabad - 221 506, India}
\end{center}
\begin{abstract}

We show the usefulness of an $e^-\g$ collider with backscattered
laser photons in unveiling the signatures of Gauge-Mediated Supersymmetry
Breaking models, with a right selectron as the next LSP. We also show how 
the signal can be distinguished from the Standard Model background 
as well as from signals of minimal supersymmetry.

\end{abstract}

\vskip 1 true cm


\newpage
\setcounter{footnote}{0}

\def\baselinestretch{1.8}
\noindent {\bf 1. Introduction}

Supersymmetry (SUSY), if exists, must be broken, and the intriguing question 
is how. Due to the fact that it cannot be broken 
in a phenomenologically consistent way in the observable sector
\cite{supertrace}, one is forced to envisage a hidden sector where SUSY
breaking takes place. How this breaking is conveyed to the observable
sector is a point of debate. The most popular theory  is that the breaking is
transmitted through  gravitational interaction, which is 
common to both the hidden  and the observable sectors \cite{sugra}. 
In this class of models, which necessarily have the gravitino
whose mass is at or above the electroweak scale, and almost invariably have
the lightest neutralino $\neu$ as the Lightest SUSY Particle (LSP),
the one with the minimal particle content is known as the
Minimal Supersymmetric Standard Model (MSSM). 
 
Another idea which has recently received considerable attention 
is that SUSY breaking is dynamically conveyed  to
the observable sector by means of a messenger sector (MS), through 
the standard model gauge interactions. To maintain gauge coupling unification,
one chooses the simplest ansatz (though by no means a necessary one
\cite{martin}) that the MS consists of chiral superfields $\Psi$ and 
$\bar{\Psi}$, which belong to a complete representation of any Grand Unification
gauge group ({\em e.g.}, $5+\overline 5$ or $10+\overline{10}$ of SU(5),
or $16+\overline{16}$ of SO(10)). The hidden
sector and the MS are coupled by means of a gauge singlet 
superfield $S$, whose scalar and auxiliary components develop vacuum 
expectation values (VEV) in the hidden sector \footnote{The scalar component 
of $S$ may, alternatively, have an explicit mass term.} This breaks 
supersymmetry in the MS through a term $\lambda S \bar{\Psi}\Psi$ in the
messenger superpotential. The quarks and leptons in the MS transform 
vectorially under the standard model (SM) gauge 
groups, and gauginos get their SUSY-breaking
masses through one-loop effects involving these MS fields. Consequently,
the sfermions get their masses at two-loop level. These class of models are 
known as Gauge-Mediated SUSY Breaking (GMSB) models \cite{gmsbmodels}.

A common feature of all the GMSB models is a light (by the standard of
terrestrial experiments) gravitino, which is invariably the LSP. Depending
on the structure of the MS, a right slepton or the lightest neutralino can
be the Next LSP (NLSP). If we demand perturbative gauge couplings upto the
unification threshold, the number of messenger generations $N_{gen}$
(of quarks and leptons) cannot be greater than four if they belong to 
the $5+\overline{5}$, and one if they are in either $10+\overline{10}$
or $16+\overline{16}$ (a generation of $10+\overline{10}$ is equivalent 
to three $5+\overline{5}$
generations). Generally speaking, the NLSP is the lightest neutralino  
if $N_{gen}=1$, and almost always a right slepton 
if $N_{gen}=$ 3 or 4. In addition to the Higgsino mass parameter
$\mu$ and the quantity $\tan \beta$ (the ratio of the two Higgs VEVs), 
the complete mass spectrum is more or less determined with
one single input, $\L$, the ratio of the VEVs of the auxiliary and the scalar 
components of the superfield $S$. The overall scale of the messenger sector
(to be designated as M here) has a relatively minor role, as it comes in only
through the renormalisation group equations giving the evolution of the 
various  parameters. 
   
The question that naturally arises is which one (if any) of the competing
schemes of SUSY breaking is more likely 
to be chosen by nature. This is more than
merely a matter of curiosity, since the breaking mechanism in turn
affects low-energy  phenomenology, based on which the various SUSY search
programmes are being carried out in collider experiments.
In a GMSB scenario, if the lightest neutralino is the NLSP,
it is almost Bino over a large region of the parameter space, and decays
into a photon (or a $Z$) and a gravitino. Thus, the signal will consist of
one or two photons in the final state, accompanied by missing energy. Such
signals in LEP-2 \cite{leptwo}, Fermilab Tevatron \cite{tevatron} and the 
Next Linear collider (NLC) \cite{nlc} have been extensively investigated.  
As there is almost no loss in branching fraction for final-state photons,
cross-sections for such processes like $e^+e^-\longrightarrow \g\g$, $e^+e^-
\longrightarrow e^+e^-\g\g$ etc.\ are much larger in GMSB than in MSSM, and
detection of GMSB is going to be straightforward.

On the other hand, if a right slepton is the NLSP, then the basic final states
arising from its decay sequence are the same as in those in MSSM. This can be
best understood by considering the case when the three right sleptons are 
practically degenerate (described in the literature as the `slepton co-NLSP
scenario' \cite{ambrosanio}). 
A pair of right selectrons produced in $e^{+}e^{-}$ collision will
then lead to $e^{+}e^{-} + {E\!\!\!\!/}$~, 
since each selectron is liable to decay 
into an electron and a gravitino. The very same final state occurs from 
a selectron pair in MSSM. It has been contended that in GMSB, the selectron 
will in general show longer tracks, due to its larger lifetime. However,
this may not be the case all over the parameter space, particularly when 
the gravitino mass is on the higher side. One, therefore, 
has to do a painstaking 
analysis of kinematic distributions such as the distribution in missing energy,
where the baffling factor is always the whole plethora of unknown parameters
in any of the models concerned. Also, in such cases, pair-production
of the lightest neutralino, which is another possible way of revealing GMSB 
signals, may be kinematically disallowed due to generally enhanced 
superparticle masses with a large number of messenger generations.
Thus it is very much desirable to look for
situations in the selectron NLSP scenario where the dominant final states
are different from those expected in MSSM. In this paper, we suggest that an
$e^-\g$ collision experiment, performed with laser backscattering in a
linear electron-positron collider, may be useful for this purpose.

The utility of an $e^-\g$ collider in probing a large area of the 
parameter space has already been pointed out \cite{cuypers}. Typically, 
a selectron together with the lightest neutralino is expected to be most
copiously produced there. The interesting thing is that, while in MSSM the
end result is a single electron with missing energy, a GMSB scenario with
a selectron NLSP leads to three electrons plus missing energy. In the rest
of the paper, we discuss some details of this signal, together with ways of
using it to our best advantage vis-a-vis SM backgrounds and residual 
MSSM effects. 

We base our discussion upon events to be seen at the NLC
(with $\sqrt{s} = 500$ GeV) running in the
$e^-\g$ mode. With a brief description of the $e^-\g$ collider in the
next section, we show our results in section 3. Section 4 concludes
the paper.

\vskip 3 true cm

\noindent{\bf 2. The {\mbox \boldmath${e^-\g}$} Collider}
 
In the $e^-\g$ collider, a low-energy but high-intensity laser beam
backscatters off the positron beam of the original $e^+e^-$ machine 
\cite{ginzburg}.
The energy of the incident photons must be so low that no multiple
scattering ({\em i.e.}, $e^+e^-$ pair production) takes place, 
whereas the intensity should be sufficiently high so that Compton-conversion
can take place with maximum efficiency. Pair production from the incident
beam decreases the conversion efficiency and can produce serious background
effects. If the energy is just below the pair-production threshold, and if
the beam is intense enough, almost all the positrons get Compton-converted
and the photon beam has almost the same luminosity as of the positron beam.
This process produces a hard collimated photon beam, with a distribution 
in the energy of the photons. The positron beam, depleted of its energy, 
is dumped.

If the initial photon beam is collinear with the $e^+$ beam, its energy 
$E_{laser}$ is bounded by
\be
{4E_{e^+}E_{laser}\over m_e^2}\ \equiv\  x\  \leq\  2(1+\sqrt{2}),
\ee
to prevent pair creation of $e^+e^-$. The photon energy spectrum ${\cal P}
(y)$, where $y=E_{\g}/E_{e^+}=\sqrt{s}_{e^-\g}/\sqrt{s}_{e^+e^-}$, is given
by \cite{cuypers, debchou}
\be
{\cal P}(y)={1\over N}\Big[1-y+{1\over 1-y}-{4y\over x(1-y)}+
{4y^2\over x^2(1-y)^2}\Big]
\ee
where
\be
N={1\over 2}+{8\over 1+x}+{7\over 2x(1+x)}+{1\over 2x(1+x)^2}
+\Big(1-{4\over x}-{8\over x^2}\Big)\ln(1+x)
\ee
normalizes $\int {\cal P}(y)dy$ to unity. In the above formulae, we take
both the $e^+$ beam and the initial laser beam to be unpolarized. In the next
section, we will present our results with a right polarized $e^-$ beam hitting
this photon beam. As we shall see, this is very useful for reducing the
SM backgrounds to the three-electron plus missing energy signal. 

\vskip 2 true cm

\noindent{\bf 3. Results}

For $e^-\g$ collisions, the lab frame and the centre-of-mass (cm) 
frame are different.
We first calculate $d\sigma/dt$ for the process $e^-\g\longrightarrow
\rslep\neu$ in the cm frame, wherein all the Mandelstam variables are 
defined: 
\bea
{d\sigma\over dt} = {1\over 16\pi s^2} & &\Big[C^2(t-\mneu^2)(t+\mrslep^2)
+{C'}^2s(s+t-\mrslep^2)\nonumber\\
& & + 2CC'(\mneu^4-t\mneu^2-{1\over 2}s\mneu^2+t\mrslep^2-\mneu^2\mrslep^2
-{1\over 2}st)\Big]
\eea
where
\be
C=\sqrt{2}ge\tan\theta_WN_{12}{1\over t-\mrslep^2},
\ee
\be
C'=\sqrt{2}ge\tan\theta_WN_{12}{1\over s}.
\ee
The above expressions were derived with a fully right-polarized $e^-$ beam.
$N_{12}$ gives the Bino component of the lightest neutralino; the other 
components, obviously, either do not contribute or are suppressed by 
$m_e/m_W$.  

The total cross-section can be obtained by integrating over $t$ and at the
same time folding with the photon energy distribution function given in
(2). The maximum possible value of $y$ is $x/(x+1)$; we set $x$, defined
in (1), to its maximum value of $4.83$. The lower limit is given by the 
kinematic threshold $(\mneu+\mrslep)^2/s_{e^+e^-}$. However, for small 
sparticle masses, we set $y_{min}=0.4$, because photons with smaller $y$ 
are scattered over such a large angle that they are effectively lost
at the interaction region. We assume BR$(\neu\rightarrow e_R\rslep) = 1/3$
and BR$(\rslep\rightarrow e_RG) = 1$, which implies that all right sleptons
are degenerate in mass. We have checked that only right sleptons can be lighter
than the lightest neutralino in the parameter space of our interest.

Here let us justify our choice of a right-polarized $e^-$ beam. The role
of beam polarization at NLC in discovering new physics has already been
highlighted \cite{polariz}. If $\mu$ is not too small, $\neu$ is 
almost always dominated by its Bino component, and its coupling to $e_R$ 
is larger due to the hypercharge of the latter. Moreover, the left-polarized
electrons do not play any role in our signal of three lepton plus missing
energy (to be precise, production of left sleptons --- if kinematically
possible --- alongwith $\neu$ will give rise to five leptons in the final
state, as can easily be seen). Thus, the possible loss in luminosity due
to a polarized beam may be compensated by its 100\% efficiency in signal
generation. Some of the possible Standard Model (SM) backgrounds that come
from one or more $W$s radiating off the electrons are also eliminated 
by using a right-polarized beam. 

The most serious SM  backgrounds come from the processes 
$e^-\g\longrightarrow e^-W^+W^-,
e^-\g Z, e^-ZZ$. The first process has a cross section of the
order of a few hundredths of a femtobarn, and is totally negligible compared
to our signal cross-section. This is because, compared to the process 
$e^-\g\longrightarrow e^-Z$, this process 
is suppressed by the $Z$-propagator as
well as by the branching fraction of two $W$s to specific lepton channels.
Serious background contributions may come from the last two processes, with
one or two soft electrons in the $t$-channel contributing to a logarithmic
enhancement. The second process may produce significant background if the
radiated photon is soft. This is circumvented by applying a softness 
cut on any two electrons, and  demanding 
that no two electrons can have energy less than 20 GeV. The third process 
can be eliminated if we put an invariant mass cut centred on the $Z$-pole
on all possible lepton pairs. However, for $M_{breaking}\leq 100$ TeV,
the signal also gets heavily suppressed. We therefore use an alternative cut,
namely, one on the opening angle of two leptons in the azimuthal plane. This is
based on the fact that for leptons produced from an on-shell $Z$, the opening
angle is almost 180$^o$. Thus, an opening angle cut at $160^o$ 
largely suppresses the background. 
We have also applied a $p_T\slash$ ~cut of 20 GeV to discriminate the signal 
from the three-lepton background without missing energy, and an angular cut 
$10^o\leq \theta_e\leq 170^o$
to ensure that no leptons are lost in beam pipe. 

The signal cross-section, with all these cuts, is plotted in Figure 1 against
the SUSY breaking scale $M$. We have taken $M/\L$=2 since the dependence of
the cross-section on this factor is not very significant (for $M/\L=10$, 
the cross-section decreases by 10\%). The reason is that all the sparticle 
masses at $M$ are determined by $\L$ alone, and only the renormalization-group
equations of the couplings contain $M$. It can be seen that the cross-section
also depends nominally on $\tan\theta$ and $\mu$. 

In Figure 2 we show the missing $E_T$ distribution for the process. From now
on, we use $\mu=300$, $\tan\beta=2$ and $M/\L=2$. This plot is shown for two
different SUSY breaking scale. Note that for lower $M$, a missing $E_T$ cut 
eats more into the signal. Energy distributions of the three leptons are
shown in Figure 3a and 3b, for $M=50$ TeV and $M=100$ TeV respectively. Here
also the peak is towards smaller energy for low $M$, as expected. In Figure
4, we show the opening angle between a pair of leptons in the azimuthal
plane for $M=100$ TeV. The distribution is rather flat, with a slight hike
towards back-to-back leptons. For $M=50$ TeV, this trend is sharper
so that the signal leptons have a distinct peak towards the
back-to-back configuration.  Thus our cuts tend to reduce the signal 
significantly for low $M$,
though the available phase space is more compared to high-$M$ case.
This feature is reflected in Figure 1.  
However, we must emphasize that even such a reduced signal can be easily
distinguished from the background, which is almost eliminated by the cuts.
Also, situations with large superparticle masses (large $\Lambda$)
are less affected by the opening angle cuts. This is precisely the region 
where the inadequacy of $e^{+}e^{-}$ machines shows up. This again
underlines the usefulness of electron-photon collisions in unveiling 
a selectron NLSP scenario.

A pertinent question is whether the above final state can still arise from 
MSSM signals. The MSSM process responsible for a similar signal is 
$e^-_R\g\longrightarrow \chi_2^0\rslep$, $\rslep\longrightarrow e\neu$,
$\chi_2^0\longrightarrow\neu \ell^+\ell^-$ ($\chi_2^0$ being the second lightst
neutralino). The angular distribution for the last process is given 
in \cite{moortgat}. 

Clearly, a gaugino-dominated $\chi_2^0$ is necessary for the above process
to be substantial, for otherwise the production is heavily suppressed. Next,
the $\chi_2^0$ has to have a large Bino component so that it can be produced 
with right-polarized electrons. In addition, the $\chi_1^0$ has to be 
gaugino-dominated, for otherwise final states with $\tau$'s will be favoured to
those with electrons. Taking all conditions into account, the tri-electron
plus missing energy signal in $e\g$ collision can be appreciable in MSSM only if
{\it (i)} $\mu >> M_{1}, M_{2}$, and {\it (ii)} $\chi^0_2$ ($\chi^0_1$)
is Bino (Wino) dominated, or alternatively {\it (iii)} if the stau and 
the selectron have a very large mass separation. Of these, {\it (ii)} 
and {\it(iii)} are difficult to
accommodate in the MSSM, especially in the one based on minimal supergravity.

\vskip 2 true cm

\noindent{\bf 4. Summary and Conclusions}

We have shown that the $e^-\g$ collider may be a very useful 
machine to find GMSB if the number of messenger generations is three (or more), 
which will make right selectron the NLSP, and, at the same time, make both
$\rslep$ and $\neu$ heavier compared to the $N_{gen}=1$ case (for a fixed 
$\L$), so that it may not be possible for an $e^+e^-$ collider to probe
both of them. We have discussed how the signal can be distinguished from 
the SM background with the help of different cuts.   
The signal  proves to be better for higher values of the parameter $\Lambda$.
For the such cases, it is easy to distinguish the signal from the conventional
MSSM backgrounds. Such a task can be somewhat difficult for low $\Lambda$ 
if both
the neutralinos in MSSM are gaugino dominated (to be precise, a Bino-dominated
$\chi_2^0$ and a Wino-dominated $\neu$ if we use a right-polarized electron
beam), or if the selectrons turn out to be much lighter than staus. 
The last mentioned situations mostly fall outside the scope of
models based on minimal supergravity. 

Though we have not discussed the case for a neutralino NLSP, it is easy to see
what happens. First of all, the neutralino pair-production cross-section
is larger than the selectron pair-production cross-section in an $e^+e^-$ 
machine, so most probably the neutralinos will be detected in the $e^+e^-$
mode itself, and can easily be identified as a GMSB signal by two-photon
final states. In the $e^-\g$ machine, the dominant final state will be 
$e^-\g\g$ plus missing energy (if $\mneu > m_Z$, a small percentage
of $\chi$ will go to $Z$ and Gravitino). With a softness cut on photon energy,
the SM background can be reduced so that one can get a clear signal. Such
a cross-check is useful even if one finds the GMSB neutralinos in the 
$e^+e^-$ machine.

{\it NOTE ADDED}: After this work had been completed, we received 
reference \cite{barger} where some related issues in $e\g$ collisions
were treated. Our study turns out to supplement theirs in a certain way,
although it had not been originally intended.

\vskip 1 true cm

\centerline{\bf Acknowledgements}

The authors thank S. Naik and S. Roy for useful discussions.

\newpage
\setcounter{page}{9}
\setlength{\unitlength}{0.240900pt}
\ifx\plotpoint\undefined\newsavebox{\plotpoint}\fi
\sbox{\plotpoint}{\rule[-0.200pt]{0.400pt}{0.400pt}}%
\begin{picture}(1500,900)(0,0)
\font\gnuplot=cmr10 at 10pt
\gnuplot
\sbox{\plotpoint}{\rule[-0.200pt]{0.400pt}{0.400pt}}%
\put(220.0,113.0){\rule[-0.200pt]{292.934pt}{0.400pt}}
\put(220.0,113.0){\rule[-0.200pt]{4.818pt}{0.400pt}}
\put(198,113){\makebox(0,0)[r]{0}}
\put(1416.0,113.0){\rule[-0.200pt]{4.818pt}{0.400pt}}
\put(220.0,222.0){\rule[-0.200pt]{4.818pt}{0.400pt}}
\put(198,222){\makebox(0,0)[r]{0.01}}
\put(1416.0,222.0){\rule[-0.200pt]{4.818pt}{0.400pt}}
\put(220.0,331.0){\rule[-0.200pt]{4.818pt}{0.400pt}}
\put(198,331){\makebox(0,0)[r]{0.02}}
\put(1416.0,331.0){\rule[-0.200pt]{4.818pt}{0.400pt}}
\put(220.0,440.0){\rule[-0.200pt]{4.818pt}{0.400pt}}
\put(198,440){\makebox(0,0)[r]{0.03}}
\put(1416.0,440.0){\rule[-0.200pt]{4.818pt}{0.400pt}}
\put(220.0,550.0){\rule[-0.200pt]{4.818pt}{0.400pt}}
\put(198,550){\makebox(0,0)[r]{0.04}}
\put(1416.0,550.0){\rule[-0.200pt]{4.818pt}{0.400pt}}
\put(220.0,659.0){\rule[-0.200pt]{4.818pt}{0.400pt}}
\put(198,659){\makebox(0,0)[r]{0.05}}
\put(1416.0,659.0){\rule[-0.200pt]{4.818pt}{0.400pt}}
\put(220.0,768.0){\rule[-0.200pt]{4.818pt}{0.400pt}}
\put(198,768){\makebox(0,0)[r]{0.06}}
\put(1416.0,768.0){\rule[-0.200pt]{4.818pt}{0.400pt}}
\put(220.0,877.0){\rule[-0.200pt]{4.818pt}{0.400pt}}
\put(198,877){\makebox(0,0)[r]{0.07}}
\put(1416.0,877.0){\rule[-0.200pt]{4.818pt}{0.400pt}}
\put(220.0,113.0){\rule[-0.200pt]{0.400pt}{4.818pt}}
\put(220,68){\makebox(0,0){20}}
\put(220.0,857.0){\rule[-0.200pt]{0.400pt}{4.818pt}}
\put(407.0,113.0){\rule[-0.200pt]{0.400pt}{4.818pt}}
\put(407,68){\makebox(0,0){40}}
\put(407.0,857.0){\rule[-0.200pt]{0.400pt}{4.818pt}}
\put(594.0,113.0){\rule[-0.200pt]{0.400pt}{4.818pt}}
\put(594,68){\makebox(0,0){60}}
\put(594.0,857.0){\rule[-0.200pt]{0.400pt}{4.818pt}}
\put(781.0,113.0){\rule[-0.200pt]{0.400pt}{4.818pt}}
\put(781,68){\makebox(0,0){80}}
\put(781.0,857.0){\rule[-0.200pt]{0.400pt}{4.818pt}}
\put(968.0,113.0){\rule[-0.200pt]{0.400pt}{4.818pt}}
\put(968,68){\makebox(0,0){100}}
\put(968.0,857.0){\rule[-0.200pt]{0.400pt}{4.818pt}}
\put(1155.0,113.0){\rule[-0.200pt]{0.400pt}{4.818pt}}
\put(1155,68){\makebox(0,0){120}}
\put(1155.0,857.0){\rule[-0.200pt]{0.400pt}{4.818pt}}
\put(1342.0,113.0){\rule[-0.200pt]{0.400pt}{4.818pt}}
\put(1342,68){\makebox(0,0){140}}
\put(1342.0,857.0){\rule[-0.200pt]{0.400pt}{4.818pt}}
\put(220.0,113.0){\rule[-0.200pt]{292.934pt}{0.400pt}}
\put(1436.0,113.0){\rule[-0.200pt]{0.400pt}{184.048pt}}
\put(220.0,877.0){\rule[-0.200pt]{292.934pt}{0.400pt}}
\put(15,495){\makebox(0,0){$\sigma$ (pb)}}
\put(828,23){\makebox(0,0){$M$ (TeV)}}
\put(220.0,113.0){\rule[-0.200pt]{0.400pt}{184.048pt}}
\sbox{\plotpoint}{\rule[-0.500pt]{1.000pt}{1.000pt}}%
\put(1306,812){\makebox(0,0)[r]{$\mu$ = 300, tan$\beta$ = 2}}
\multiput(1328,812)(20.756,0.000){4}{\usebox{\plotpoint}}
\put(1394,812){\usebox{\plotpoint}}
\put(360,424){\usebox{\plotpoint}}
\multiput(360,424)(9.525,18.441){5}{\usebox{\plotpoint}}
\multiput(407,515)(12.411,16.636){4}{\usebox{\plotpoint}}
\multiput(454,578)(13.917,15.398){4}{\usebox{\plotpoint}}
\multiput(501,630)(15.831,13.422){3}{\usebox{\plotpoint}}
\multiput(547,669)(17.664,10.899){2}{\usebox{\plotpoint}}
\multiput(594,698)(19.518,7.060){3}{\usebox{\plotpoint}}
\multiput(641,715)(20.681,1.760){2}{\usebox{\plotpoint}}
\multiput(688,719)(20.282,-4.409){2}{\usebox{\plotpoint}}
\multiput(734,709)(18.643,-9.123){3}{\usebox{\plotpoint}}
\multiput(781,686)(16.140,-13.049){3}{\usebox{\plotpoint}}
\multiput(828,648)(14.066,-15.263){3}{\usebox{\plotpoint}}
\multiput(875,597)(11.236,-17.451){4}{\usebox{\plotpoint}}
\multiput(922,524)(11.065,-17.560){4}{\usebox{\plotpoint}}
\multiput(968,451)(11.801,-17.074){4}{\usebox{\plotpoint}}
\multiput(1015,383)(12.539,-16.540){4}{\usebox{\plotpoint}}
\multiput(1062,321)(13.343,-15.898){3}{\usebox{\plotpoint}}
\multiput(1109,265)(14.361,-14.985){4}{\usebox{\plotpoint}}
\multiput(1155,217)(15.806,-13.452){3}{\usebox{\plotpoint}}
\multiput(1202,177)(17.156,-11.681){2}{\usebox{\plotpoint}}
\multiput(1249,145)(18.798,-8.799){3}{\usebox{\plotpoint}}
\multiput(1296,123)(20.282,-4.409){2}{\usebox{\plotpoint}}
\multiput(1342,113)(20.756,0.000){2}{\usebox{\plotpoint}}
\multiput(1389,113)(20.756,0.000){3}{\usebox{\plotpoint}}
\put(1436,113){\usebox{\plotpoint}}
\sbox{\plotpoint}{\rule[-0.200pt]{0.400pt}{0.400pt}}%
\put(1306,767){\makebox(0,0)[r]{$\mu$ = 300, tan$\beta$ = 40}}
\put(1328.0,767.0){\rule[-0.200pt]{15.899pt}{0.400pt}}
\put(314,419){\usebox{\plotpoint}}
\multiput(314.58,419.00)(0.498,0.937){89}{\rule{0.120pt}{0.848pt}}
\multiput(313.17,419.00)(46.000,84.240){2}{\rule{0.400pt}{0.424pt}}
\multiput(360.58,505.00)(0.498,0.874){91}{\rule{0.120pt}{0.798pt}}
\multiput(359.17,505.00)(47.000,80.344){2}{\rule{0.400pt}{0.399pt}}
\multiput(407.58,587.00)(0.498,0.628){91}{\rule{0.120pt}{0.602pt}}
\multiput(406.17,587.00)(47.000,57.750){2}{\rule{0.400pt}{0.301pt}}
\multiput(454.00,646.58)(0.510,0.498){89}{\rule{0.509pt}{0.120pt}}
\multiput(454.00,645.17)(45.944,46.000){2}{\rule{0.254pt}{0.400pt}}
\multiput(501.00,692.58)(0.622,0.498){71}{\rule{0.597pt}{0.120pt}}
\multiput(501.00,691.17)(44.760,37.000){2}{\rule{0.299pt}{0.400pt}}
\multiput(547.00,729.58)(0.945,0.497){47}{\rule{0.852pt}{0.120pt}}
\multiput(547.00,728.17)(45.232,25.000){2}{\rule{0.426pt}{0.400pt}}
\multiput(594.00,754.58)(2.194,0.492){19}{\rule{1.809pt}{0.118pt}}
\multiput(594.00,753.17)(43.245,11.000){2}{\rule{0.905pt}{0.400pt}}
\put(641,763.17){\rule{9.500pt}{0.400pt}}
\multiput(641.00,764.17)(27.282,-2.000){2}{\rule{4.750pt}{0.400pt}}
\multiput(688.00,761.92)(1.458,-0.494){29}{\rule{1.250pt}{0.119pt}}
\multiput(688.00,762.17)(43.406,-16.000){2}{\rule{0.625pt}{0.400pt}}
\multiput(734.00,745.92)(0.760,-0.497){59}{\rule{0.706pt}{0.120pt}}
\multiput(734.00,746.17)(45.534,-31.000){2}{\rule{0.353pt}{0.400pt}}
\multiput(781.58,713.89)(0.498,-0.510){91}{\rule{0.120pt}{0.509pt}}
\multiput(780.17,714.94)(47.000,-46.945){2}{\rule{0.400pt}{0.254pt}}
\multiput(828.58,665.08)(0.498,-0.756){91}{\rule{0.120pt}{0.704pt}}
\multiput(827.17,666.54)(47.000,-69.538){2}{\rule{0.400pt}{0.352pt}}
\multiput(875.58,593.76)(0.498,-0.853){91}{\rule{0.120pt}{0.781pt}}
\multiput(874.17,595.38)(47.000,-78.379){2}{\rule{0.400pt}{0.390pt}}
\multiput(922.58,513.81)(0.498,-0.839){89}{\rule{0.120pt}{0.770pt}}
\multiput(921.17,515.40)(46.000,-75.403){2}{\rule{0.400pt}{0.385pt}}
\multiput(968.58,437.04)(0.498,-0.767){91}{\rule{0.120pt}{0.713pt}}
\multiput(967.17,438.52)(47.000,-70.521){2}{\rule{0.400pt}{0.356pt}}
\multiput(1015.58,365.25)(0.498,-0.703){91}{\rule{0.120pt}{0.662pt}}
\multiput(1014.17,366.63)(47.000,-64.627){2}{\rule{0.400pt}{0.331pt}}
\multiput(1062.58,299.47)(0.498,-0.639){91}{\rule{0.120pt}{0.611pt}}
\multiput(1061.17,300.73)(47.000,-58.733){2}{\rule{0.400pt}{0.305pt}}
\multiput(1109.58,239.74)(0.498,-0.554){89}{\rule{0.120pt}{0.543pt}}
\multiput(1108.17,240.87)(46.000,-49.872){2}{\rule{0.400pt}{0.272pt}}
\multiput(1155.00,189.92)(0.573,-0.498){79}{\rule{0.559pt}{0.120pt}}
\multiput(1155.00,190.17)(45.841,-41.000){2}{\rule{0.279pt}{0.400pt}}
\multiput(1202.00,148.92)(0.813,-0.497){55}{\rule{0.748pt}{0.120pt}}
\multiput(1202.00,149.17)(45.447,-29.000){2}{\rule{0.374pt}{0.400pt}}
\multiput(1249.00,119.93)(3.069,-0.488){13}{\rule{2.450pt}{0.117pt}}
\multiput(1249.00,120.17)(41.915,-8.000){2}{\rule{1.225pt}{0.400pt}}
\put(1296.0,113.0){\rule[-0.200pt]{33.726pt}{0.400pt}}
\sbox{\plotpoint}{\rule[-0.400pt]{0.800pt}{0.800pt}}%
\put(1306,722){\makebox(0,0)[r]{$\mu = -900$, tan$\beta$ = 40}}
\put(1328.0,722.0){\rule[-0.400pt]{15.899pt}{0.800pt}}
\put(314,434){\usebox{\plotpoint}}
\multiput(315.41,434.00)(0.502,1.227){85}{\rule{0.121pt}{2.148pt}}
\multiput(312.34,434.00)(46.000,107.542){2}{\rule{0.800pt}{1.074pt}}
\multiput(361.41,546.00)(0.502,0.801){87}{\rule{0.121pt}{1.477pt}}
\multiput(358.34,546.00)(47.000,71.935){2}{\rule{0.800pt}{0.738pt}}
\multiput(408.41,621.00)(0.502,0.650){87}{\rule{0.121pt}{1.238pt}}
\multiput(405.34,621.00)(47.000,58.430){2}{\rule{0.800pt}{0.619pt}}
\multiput(455.41,682.00)(0.502,0.520){87}{\rule{0.121pt}{1.034pt}}
\multiput(452.34,682.00)(47.000,46.854){2}{\rule{0.800pt}{0.517pt}}
\multiput(501.00,732.41)(0.574,0.502){73}{\rule{1.120pt}{0.121pt}}
\multiput(501.00,729.34)(43.675,40.000){2}{\rule{0.560pt}{0.800pt}}
\multiput(547.00,772.41)(0.913,0.504){45}{\rule{1.646pt}{0.121pt}}
\multiput(547.00,769.34)(43.583,26.000){2}{\rule{0.823pt}{0.800pt}}
\multiput(594.00,798.41)(1.751,0.509){21}{\rule{2.886pt}{0.123pt}}
\multiput(594.00,795.34)(41.011,14.000){2}{\rule{1.443pt}{0.800pt}}
\put(641,808.84){\rule{11.322pt}{0.800pt}}
\multiput(641.00,809.34)(23.500,-1.000){2}{\rule{5.661pt}{0.800pt}}
\multiput(688.00,808.09)(1.310,-0.506){29}{\rule{2.244pt}{0.122pt}}
\multiput(688.00,808.34)(41.342,-18.000){2}{\rule{1.122pt}{0.800pt}}
\multiput(734.00,790.09)(0.693,-0.503){61}{\rule{1.306pt}{0.121pt}}
\multiput(734.00,790.34)(44.290,-34.000){2}{\rule{0.653pt}{0.800pt}}
\multiput(782.41,753.28)(0.502,-0.585){87}{\rule{0.121pt}{1.136pt}}
\multiput(779.34,755.64)(47.000,-52.642){2}{\rule{0.800pt}{0.568pt}}
\multiput(829.41,696.31)(0.502,-0.887){87}{\rule{0.121pt}{1.613pt}}
\multiput(826.34,699.65)(47.000,-79.653){2}{\rule{0.800pt}{0.806pt}}
\multiput(876.41,613.16)(0.502,-0.909){87}{\rule{0.121pt}{1.647pt}}
\multiput(873.34,616.58)(47.000,-81.582){2}{\rule{0.800pt}{0.823pt}}
\multiput(923.41,528.18)(0.502,-0.907){85}{\rule{0.121pt}{1.643pt}}
\multiput(920.34,531.59)(46.000,-79.589){2}{\rule{0.800pt}{0.822pt}}
\multiput(969.41,445.59)(0.502,-0.844){87}{\rule{0.121pt}{1.545pt}}
\multiput(966.34,448.79)(47.000,-75.794){2}{\rule{0.800pt}{0.772pt}}
\multiput(1016.41,366.80)(0.502,-0.812){87}{\rule{0.121pt}{1.494pt}}
\multiput(1013.34,369.90)(47.000,-72.900){2}{\rule{0.800pt}{0.747pt}}
\multiput(1063.41,291.15)(0.502,-0.758){87}{\rule{0.121pt}{1.409pt}}
\multiput(1060.34,294.08)(47.000,-68.077){2}{\rule{0.800pt}{0.704pt}}
\multiput(1110.41,220.55)(0.502,-0.697){85}{\rule{0.121pt}{1.313pt}}
\multiput(1107.34,223.27)(46.000,-61.275){2}{\rule{0.800pt}{0.657pt}}
\multiput(1155.00,160.09)(0.498,-0.502){87}{\rule{1.000pt}{0.121pt}}
\multiput(1155.00,160.34)(44.924,-47.000){2}{\rule{0.500pt}{0.800pt}}
\put(1202,112.34){\rule{11.322pt}{0.800pt}}
\multiput(1202.00,113.34)(23.500,-2.000){2}{\rule{5.661pt}{0.800pt}}
\put(1249.0,113.0){\rule[-0.400pt]{45.048pt}{0.800pt}}
\sbox{\plotpoint}{\rule[-0.600pt]{1.200pt}{1.200pt}}%
\put(1306,677){\makebox(0,0)[r]{$\mu = -900$, tan$\beta$ = 2}}
\put(1328.0,677.0){\rule[-0.600pt]{15.899pt}{1.200pt}}
\put(360,566){\usebox{\plotpoint}}
\multiput(362.24,566.00)(0.500,0.728){84}{\rule{0.121pt}{2.062pt}}
\multiput(357.51,566.00)(47.000,64.721){2}{\rule{1.200pt}{1.031pt}}
\multiput(409.24,635.00)(0.500,0.642){84}{\rule{0.121pt}{1.857pt}}
\multiput(404.51,635.00)(47.000,57.145){2}{\rule{1.200pt}{0.929pt}}
\multiput(456.24,696.00)(0.500,0.545){84}{\rule{0.121pt}{1.628pt}}
\multiput(451.51,696.00)(47.000,48.622){2}{\rule{1.200pt}{0.814pt}}
\multiput(501.00,750.24)(0.566,0.500){70}{\rule{1.680pt}{0.121pt}}
\multiput(501.00,745.51)(42.513,40.000){2}{\rule{0.840pt}{1.200pt}}
\multiput(547.00,790.24)(0.831,0.500){46}{\rule{2.314pt}{0.121pt}}
\multiput(547.00,785.51)(42.197,28.000){2}{\rule{1.157pt}{1.200pt}}
\multiput(594.00,818.24)(1.708,0.501){18}{\rule{4.329pt}{0.121pt}}
\multiput(594.00,813.51)(38.016,14.000){2}{\rule{2.164pt}{1.200pt}}
\put(641,826.01){\rule{11.322pt}{1.200pt}}
\multiput(641.00,827.51)(23.500,-3.000){2}{\rule{5.661pt}{1.200pt}}
\multiput(688.00,824.26)(1.281,-0.501){26}{\rule{3.367pt}{0.121pt}}
\multiput(688.00,824.51)(39.012,-18.000){2}{\rule{1.683pt}{1.200pt}}
\multiput(734.00,806.26)(0.643,-0.500){62}{\rule{1.867pt}{0.121pt}}
\multiput(734.00,806.51)(43.126,-36.000){2}{\rule{0.933pt}{1.200pt}}
\multiput(783.24,765.93)(0.500,-0.577){84}{\rule{0.121pt}{1.704pt}}
\multiput(778.51,769.46)(47.000,-51.463){2}{\rule{1.200pt}{0.852pt}}
\multiput(830.24,707.64)(0.500,-0.911){84}{\rule{0.121pt}{2.496pt}}
\multiput(825.51,712.82)(47.000,-80.820){2}{\rule{1.200pt}{1.248pt}}
\multiput(877.24,621.53)(0.500,-0.922){84}{\rule{0.121pt}{2.521pt}}
\multiput(872.51,626.77)(47.000,-81.767){2}{\rule{1.200pt}{1.261pt}}
\multiput(924.24,534.55)(0.500,-0.920){82}{\rule{0.121pt}{2.517pt}}
\multiput(919.51,539.78)(46.000,-79.775){2}{\rule{1.200pt}{1.259pt}}
\multiput(970.24,450.06)(0.500,-0.868){84}{\rule{0.121pt}{2.394pt}}
\multiput(965.51,455.03)(47.000,-77.032){2}{\rule{1.200pt}{1.197pt}}
\multiput(1017.24,368.59)(0.500,-0.814){84}{\rule{0.121pt}{2.266pt}}
\multiput(1012.51,373.30)(47.000,-72.297){2}{\rule{1.200pt}{1.133pt}}
\multiput(1064.24,292.12)(0.500,-0.761){84}{\rule{0.121pt}{2.138pt}}
\multiput(1059.51,296.56)(47.000,-67.562){2}{\rule{1.200pt}{1.069pt}}
\multiput(1111.24,220.72)(0.500,-0.700){82}{\rule{0.121pt}{1.996pt}}
\multiput(1106.51,224.86)(46.000,-60.858){2}{\rule{1.200pt}{0.998pt}}
\multiput(1157.24,157.56)(0.500,-0.513){84}{\rule{0.121pt}{1.551pt}}
\multiput(1152.51,160.78)(47.000,-45.781){2}{\rule{1.200pt}{0.776pt}}
\put(1202,111.51){\rule{11.322pt}{1.200pt}}
\multiput(1202.00,112.51)(23.500,-2.000){2}{\rule{5.661pt}{1.200pt}}
\put(1249.0,113.0){\rule[-0.600pt]{45.048pt}{1.200pt}}
\end{picture}
\vskip 1 in
\begin{center}
\bf{Figure 1}
\end{center}
\vskip .1 in
\noindent
Cross-section for the process $e^-\gamma\longrightarrow e^-e^+e^- + E
{\!\!\!\!/}$~~, plotted against the SUSY breaking scale $M$. The cuts 
are as discussed in the text, and M/$\Lambda$ = 2   
\newpage
\setlength{\unitlength}{0.240900pt}
\ifx\plotpoint\undefined\newsavebox{\plotpoint}\fi
\sbox{\plotpoint}{\rule[-0.200pt]{0.400pt}{0.400pt}}%
\begin{picture}(1500,900)(0,0)
\font\gnuplot=cmr10 at 10pt
\gnuplot
\sbox{\plotpoint}{\rule[-0.200pt]{0.400pt}{0.400pt}}%
\put(220.0,113.0){\rule[-0.200pt]{292.934pt}{0.400pt}}
\put(220.0,113.0){\rule[-0.200pt]{0.400pt}{184.048pt}}
\put(220.0,113.0){\rule[-0.200pt]{4.818pt}{0.400pt}}
\put(198,113){\makebox(0,0)[r]{0}}
\put(1416.0,113.0){\rule[-0.200pt]{4.818pt}{0.400pt}}
\put(220.0,222.0){\rule[-0.200pt]{4.818pt}{0.400pt}}
\put(198,222){\makebox(0,0)[r]{0.0005}}
\put(1416.0,222.0){\rule[-0.200pt]{4.818pt}{0.400pt}}
\put(220.0,331.0){\rule[-0.200pt]{4.818pt}{0.400pt}}
\put(198,331){\makebox(0,0)[r]{0.001}}
\put(1416.0,331.0){\rule[-0.200pt]{4.818pt}{0.400pt}}
\put(220.0,440.0){\rule[-0.200pt]{4.818pt}{0.400pt}}
\put(198,440){\makebox(0,0)[r]{0.0015}}
\put(1416.0,440.0){\rule[-0.200pt]{4.818pt}{0.400pt}}
\put(220.0,550.0){\rule[-0.200pt]{4.818pt}{0.400pt}}
\put(198,550){\makebox(0,0)[r]{0.002}}
\put(1416.0,550.0){\rule[-0.200pt]{4.818pt}{0.400pt}}
\put(220.0,659.0){\rule[-0.200pt]{4.818pt}{0.400pt}}
\put(198,659){\makebox(0,0)[r]{0.0025}}
\put(1416.0,659.0){\rule[-0.200pt]{4.818pt}{0.400pt}}
\put(220.0,768.0){\rule[-0.200pt]{4.818pt}{0.400pt}}
\put(198,768){\makebox(0,0)[r]{0.003}}
\put(1416.0,768.0){\rule[-0.200pt]{4.818pt}{0.400pt}}
\put(220.0,877.0){\rule[-0.200pt]{4.818pt}{0.400pt}}
\put(198,877){\makebox(0,0)[r]{0.0035}}
\put(1416.0,877.0){\rule[-0.200pt]{4.818pt}{0.400pt}}
\put(220.0,113.0){\rule[-0.200pt]{0.400pt}{4.818pt}}
\put(220,68){\makebox(0,0){0}}
\put(220.0,857.0){\rule[-0.200pt]{0.400pt}{4.818pt}}
\put(463.0,113.0){\rule[-0.200pt]{0.400pt}{4.818pt}}
\put(463,68){\makebox(0,0){50}}
\put(463.0,857.0){\rule[-0.200pt]{0.400pt}{4.818pt}}
\put(706.0,113.0){\rule[-0.200pt]{0.400pt}{4.818pt}}
\put(706,68){\makebox(0,0){100}}
\put(706.0,857.0){\rule[-0.200pt]{0.400pt}{4.818pt}}
\put(950.0,113.0){\rule[-0.200pt]{0.400pt}{4.818pt}}
\put(950,68){\makebox(0,0){150}}
\put(950.0,857.0){\rule[-0.200pt]{0.400pt}{4.818pt}}
\put(1193.0,113.0){\rule[-0.200pt]{0.400pt}{4.818pt}}
\put(1193,68){\makebox(0,0){200}}
\put(1193.0,857.0){\rule[-0.200pt]{0.400pt}{4.818pt}}
\put(1436.0,113.0){\rule[-0.200pt]{0.400pt}{4.818pt}}
\put(1436,68){\makebox(0,0){250}}
\put(1436.0,857.0){\rule[-0.200pt]{0.400pt}{4.818pt}}
\put(220.0,113.0){\rule[-0.200pt]{292.934pt}{0.400pt}}
\put(1436.0,113.0){\rule[-0.200pt]{0.400pt}{184.048pt}}
\put(220.0,877.0){\rule[-0.200pt]{292.934pt}{0.400pt}}
\put(15,495){\makebox(0,0){$d\sigma$ (pb)}}
\put(828,13){\makebox(0,0){$E{\!\!\!\!/_T}$~~ (GeV)}}
\put(220.0,113.0){\rule[-0.200pt]{0.400pt}{184.048pt}}
\put(1306,812){\makebox(0,0)[r]{$M$ = 50 TeV}}
\put(1328.0,812.0){\rule[-0.200pt]{15.899pt}{0.400pt}}
\put(220,136){\usebox{\plotpoint}}
\put(220.0,136.0){\rule[-0.200pt]{5.782pt}{0.400pt}}
\put(244.0,136.0){\rule[-0.200pt]{0.400pt}{11.563pt}}
\put(244.0,184.0){\rule[-0.200pt]{6.022pt}{0.400pt}}
\put(269.0,184.0){\rule[-0.200pt]{0.400pt}{7.468pt}}
\put(269.0,215.0){\rule[-0.200pt]{5.782pt}{0.400pt}}
\put(293.0,215.0){\rule[-0.200pt]{0.400pt}{10.359pt}}
\put(293.0,258.0){\rule[-0.200pt]{5.782pt}{0.400pt}}
\put(317.0,258.0){\rule[-0.200pt]{0.400pt}{11.804pt}}
\put(317.0,307.0){\rule[-0.200pt]{6.022pt}{0.400pt}}
\put(342.0,307.0){\rule[-0.200pt]{0.400pt}{13.972pt}}
\put(342.0,365.0){\rule[-0.200pt]{5.782pt}{0.400pt}}
\put(366.0,365.0){\rule[-0.200pt]{0.400pt}{14.936pt}}
\put(366.0,427.0){\rule[-0.200pt]{5.782pt}{0.400pt}}
\put(390.0,427.0){\rule[-0.200pt]{0.400pt}{17.104pt}}
\put(390.0,498.0){\rule[-0.200pt]{6.022pt}{0.400pt}}
\put(415.0,498.0){\rule[-0.200pt]{0.400pt}{20.717pt}}
\put(415.0,584.0){\rule[-0.200pt]{5.782pt}{0.400pt}}
\put(439.0,584.0){\rule[-0.200pt]{0.400pt}{17.827pt}}
\put(439.0,658.0){\rule[-0.200pt]{5.782pt}{0.400pt}}
\put(463.0,658.0){\rule[-0.200pt]{0.400pt}{12.045pt}}
\put(463.0,708.0){\rule[-0.200pt]{6.022pt}{0.400pt}}
\put(488.0,708.0){\rule[-0.200pt]{0.400pt}{14.936pt}}
\put(488.0,770.0){\rule[-0.200pt]{5.782pt}{0.400pt}}
\put(512.0,770.0){\rule[-0.200pt]{0.400pt}{6.263pt}}
\put(512.0,796.0){\rule[-0.200pt]{5.782pt}{0.400pt}}
\put(536.0,796.0){\rule[-0.200pt]{0.400pt}{4.336pt}}
\put(536.0,814.0){\rule[-0.200pt]{5.782pt}{0.400pt}}
\put(560.0,804.0){\rule[-0.200pt]{0.400pt}{2.409pt}}
\put(560.0,804.0){\rule[-0.200pt]{6.022pt}{0.400pt}}
\put(585.0,754.0){\rule[-0.200pt]{0.400pt}{12.045pt}}
\put(585.0,754.0){\rule[-0.200pt]{5.782pt}{0.400pt}}
\put(609.0,683.0){\rule[-0.200pt]{0.400pt}{17.104pt}}
\put(609.0,683.0){\rule[-0.200pt]{5.782pt}{0.400pt}}
\put(633.0,585.0){\rule[-0.200pt]{0.400pt}{23.608pt}}
\put(633.0,585.0){\rule[-0.200pt]{6.022pt}{0.400pt}}
\put(658.0,526.0){\rule[-0.200pt]{0.400pt}{14.213pt}}
\put(658.0,526.0){\rule[-0.200pt]{5.782pt}{0.400pt}}
\put(682.0,484.0){\rule[-0.200pt]{0.400pt}{10.118pt}}
\put(682.0,484.0){\rule[-0.200pt]{5.782pt}{0.400pt}}
\put(706.0,456.0){\rule[-0.200pt]{0.400pt}{6.745pt}}
\put(706.0,456.0){\rule[-0.200pt]{6.022pt}{0.400pt}}
\put(731.0,404.0){\rule[-0.200pt]{0.400pt}{12.527pt}}
\put(731.0,404.0){\rule[-0.200pt]{5.782pt}{0.400pt}}
\put(755.0,374.0){\rule[-0.200pt]{0.400pt}{7.227pt}}
\put(755.0,374.0){\rule[-0.200pt]{5.782pt}{0.400pt}}
\put(779.0,347.0){\rule[-0.200pt]{0.400pt}{6.504pt}}
\put(779.0,347.0){\rule[-0.200pt]{6.022pt}{0.400pt}}
\put(804.0,323.0){\rule[-0.200pt]{0.400pt}{5.782pt}}
\put(804.0,323.0){\rule[-0.200pt]{5.782pt}{0.400pt}}
\put(828.0,290.0){\rule[-0.200pt]{0.400pt}{7.950pt}}
\put(828.0,290.0){\rule[-0.200pt]{5.782pt}{0.400pt}}
\put(852.0,269.0){\rule[-0.200pt]{0.400pt}{5.059pt}}
\put(852.0,269.0){\rule[-0.200pt]{6.022pt}{0.400pt}}
\put(877.0,250.0){\rule[-0.200pt]{0.400pt}{4.577pt}}
\put(877.0,250.0){\rule[-0.200pt]{5.782pt}{0.400pt}}
\put(901.0,222.0){\rule[-0.200pt]{0.400pt}{6.745pt}}
\put(901.0,222.0){\rule[-0.200pt]{5.782pt}{0.400pt}}
\put(925.0,212.0){\rule[-0.200pt]{0.400pt}{2.409pt}}
\put(925.0,212.0){\rule[-0.200pt]{6.022pt}{0.400pt}}
\put(950.0,196.0){\rule[-0.200pt]{0.400pt}{3.854pt}}
\put(950.0,196.0){\rule[-0.200pt]{5.782pt}{0.400pt}}
\put(974.0,180.0){\rule[-0.200pt]{0.400pt}{3.854pt}}
\put(974.0,180.0){\rule[-0.200pt]{5.782pt}{0.400pt}}
\put(998.0,168.0){\rule[-0.200pt]{0.400pt}{2.891pt}}
\put(998.0,168.0){\rule[-0.200pt]{6.022pt}{0.400pt}}
\put(1023.0,156.0){\rule[-0.200pt]{0.400pt}{2.891pt}}
\put(1023.0,156.0){\rule[-0.200pt]{5.782pt}{0.400pt}}
\put(1047.0,148.0){\rule[-0.200pt]{0.400pt}{1.927pt}}
\put(1047.0,148.0){\rule[-0.200pt]{5.782pt}{0.400pt}}
\put(1071.0,139.0){\rule[-0.200pt]{0.400pt}{2.168pt}}
\put(1071.0,139.0){\rule[-0.200pt]{6.022pt}{0.400pt}}
\put(1096.0,134.0){\rule[-0.200pt]{0.400pt}{1.204pt}}
\put(1096.0,134.0){\rule[-0.200pt]{5.782pt}{0.400pt}}
\put(1120.0,127.0){\rule[-0.200pt]{0.400pt}{1.686pt}}
\put(1120.0,127.0){\rule[-0.200pt]{5.782pt}{0.400pt}}
\put(1144.0,123.0){\rule[-0.200pt]{0.400pt}{0.964pt}}
\put(1144.0,123.0){\rule[-0.200pt]{5.782pt}{0.400pt}}
\put(1168.0,120.0){\rule[-0.200pt]{0.400pt}{0.723pt}}
\put(1168.0,120.0){\rule[-0.200pt]{6.022pt}{0.400pt}}
\put(1193.0,118.0){\rule[-0.200pt]{0.400pt}{0.482pt}}
\put(1193.0,118.0){\rule[-0.200pt]{5.782pt}{0.400pt}}
\put(1217.0,115.0){\rule[-0.200pt]{0.400pt}{0.723pt}}
\put(1217.0,115.0){\rule[-0.200pt]{5.782pt}{0.400pt}}
\put(1241.0,114.0){\usebox{\plotpoint}}
\put(1241.0,114.0){\rule[-0.200pt]{6.022pt}{0.400pt}}
\put(1266.0,113.0){\usebox{\plotpoint}}
\put(1266.0,113.0){\rule[-0.200pt]{40.953pt}{0.400pt}}
\sbox{\plotpoint}{\rule[-0.400pt]{0.800pt}{0.800pt}}%
\put(1306,767){\makebox(0,0)[r]{$M$ = 100 TeV}}
\put(1328.0,767.0){\rule[-0.400pt]{15.899pt}{0.800pt}}
\put(220,125){\usebox{\plotpoint}}
\put(220.0,125.0){\rule[-0.400pt]{5.782pt}{0.800pt}}
\put(244.0,125.0){\rule[-0.400pt]{0.800pt}{5.541pt}}
\put(244.0,148.0){\rule[-0.400pt]{6.022pt}{0.800pt}}
\put(269.0,148.0){\rule[-0.400pt]{0.800pt}{4.336pt}}
\put(269.0,166.0){\rule[-0.400pt]{5.782pt}{0.800pt}}
\put(293.0,166.0){\rule[-0.400pt]{0.800pt}{3.373pt}}
\put(293.0,180.0){\rule[-0.400pt]{5.782pt}{0.800pt}}
\put(317.0,180.0){\rule[-0.400pt]{0.800pt}{3.613pt}}
\put(317.0,195.0){\rule[-0.400pt]{6.022pt}{0.800pt}}
\put(342.0,195.0){\rule[-0.400pt]{0.800pt}{1.927pt}}
\put(342.0,203.0){\rule[-0.400pt]{5.782pt}{0.800pt}}
\put(366.0,203.0){\rule[-0.400pt]{0.800pt}{4.336pt}}
\put(366.0,221.0){\rule[-0.400pt]{5.782pt}{0.800pt}}
\put(390.0,221.0){\rule[-0.400pt]{0.800pt}{1.927pt}}
\put(390.0,229.0){\rule[-0.400pt]{6.022pt}{0.800pt}}
\put(415.0,229.0){\rule[-0.400pt]{0.800pt}{3.854pt}}
\put(415.0,245.0){\rule[-0.400pt]{5.782pt}{0.800pt}}
\put(439.0,245.0){\rule[-0.400pt]{0.800pt}{5.782pt}}
\put(439.0,269.0){\rule[-0.400pt]{5.782pt}{0.800pt}}
\put(463.0,269.0){\rule[-0.400pt]{0.800pt}{3.613pt}}
\put(463.0,284.0){\rule[-0.400pt]{6.022pt}{0.800pt}}
\put(488.0,284.0){\rule[-0.400pt]{0.800pt}{5.541pt}}
\put(488.0,307.0){\rule[-0.400pt]{5.782pt}{0.800pt}}
\put(512.0,307.0){\rule[-0.400pt]{0.800pt}{5.541pt}}
\put(512.0,330.0){\rule[-0.400pt]{5.782pt}{0.800pt}}
\put(536.0,330.0){\rule[-0.400pt]{0.800pt}{6.022pt}}
\put(536.0,355.0){\rule[-0.400pt]{5.782pt}{0.800pt}}
\put(560.0,355.0){\rule[-0.400pt]{0.800pt}{3.132pt}}
\put(560.0,368.0){\rule[-0.400pt]{6.022pt}{0.800pt}}
\put(585.0,368.0){\rule[-0.400pt]{0.800pt}{4.095pt}}
\put(585.0,385.0){\rule[-0.400pt]{5.782pt}{0.800pt}}
\put(609.0,385.0){\rule[-0.400pt]{0.800pt}{3.373pt}}
\put(609.0,399.0){\rule[-0.400pt]{5.782pt}{0.800pt}}
\put(633.0,399.0){\rule[-0.400pt]{0.800pt}{2.650pt}}
\put(633.0,410.0){\rule[-0.400pt]{6.022pt}{0.800pt}}
\put(658.0,410.0){\rule[-0.400pt]{0.800pt}{1.445pt}}
\put(658.0,416.0){\rule[-0.400pt]{5.782pt}{0.800pt}}
\put(682.0,416.0){\rule[-0.400pt]{0.800pt}{2.650pt}}
\put(682.0,427.0){\rule[-0.400pt]{5.782pt}{0.800pt}}
\put(706.0,427.0){\usebox{\plotpoint}}
\put(706.0,429.0){\rule[-0.400pt]{6.022pt}{0.800pt}}
\put(731.0,427.0){\usebox{\plotpoint}}
\put(731.0,427.0){\rule[-0.400pt]{5.782pt}{0.800pt}}
\put(755.0,427.0){\rule[-0.400pt]{0.800pt}{0.964pt}}
\put(755.0,431.0){\rule[-0.400pt]{5.782pt}{0.800pt}}
\put(779.0,425.0){\rule[-0.400pt]{0.800pt}{1.445pt}}
\put(779.0,425.0){\rule[-0.400pt]{6.022pt}{0.800pt}}
\put(804.0,414.0){\rule[-0.400pt]{0.800pt}{2.650pt}}
\put(804.0,414.0){\rule[-0.400pt]{5.782pt}{0.800pt}}
\put(828.0,402.0){\rule[-0.400pt]{0.800pt}{2.891pt}}
\put(828.0,402.0){\rule[-0.400pt]{5.782pt}{0.800pt}}
\put(852.0,398.0){\rule[-0.400pt]{0.800pt}{0.964pt}}
\put(852.0,398.0){\rule[-0.400pt]{6.022pt}{0.800pt}}
\put(877.0,372.0){\rule[-0.400pt]{0.800pt}{6.263pt}}
\put(877.0,372.0){\rule[-0.400pt]{5.782pt}{0.800pt}}
\put(901.0,347.0){\rule[-0.400pt]{0.800pt}{6.022pt}}
\put(901.0,347.0){\rule[-0.400pt]{5.782pt}{0.800pt}}
\put(925.0,321.0){\rule[-0.400pt]{0.800pt}{6.263pt}}
\put(925.0,321.0){\rule[-0.400pt]{6.022pt}{0.800pt}}
\put(950.0,291.0){\rule[-0.400pt]{0.800pt}{7.227pt}}
\put(950.0,291.0){\rule[-0.400pt]{5.782pt}{0.800pt}}
\put(974.0,255.0){\rule[-0.400pt]{0.800pt}{8.672pt}}
\put(974.0,255.0){\rule[-0.400pt]{5.782pt}{0.800pt}}
\put(998.0,223.0){\rule[-0.400pt]{0.800pt}{7.709pt}}
\put(998.0,223.0){\rule[-0.400pt]{6.022pt}{0.800pt}}
\put(1023.0,193.0){\rule[-0.400pt]{0.800pt}{7.227pt}}
\put(1023.0,193.0){\rule[-0.400pt]{5.782pt}{0.800pt}}
\put(1047.0,172.0){\rule[-0.400pt]{0.800pt}{5.059pt}}
\put(1047.0,172.0){\rule[-0.400pt]{5.782pt}{0.800pt}}
\put(1071.0,154.0){\rule[-0.400pt]{0.800pt}{4.336pt}}
\put(1071.0,154.0){\rule[-0.400pt]{6.022pt}{0.800pt}}
\put(1096.0,142.0){\rule[-0.400pt]{0.800pt}{2.891pt}}
\put(1096.0,142.0){\rule[-0.400pt]{5.782pt}{0.800pt}}
\put(1120.0,134.0){\rule[-0.400pt]{0.800pt}{1.927pt}}
\put(1120.0,134.0){\rule[-0.400pt]{5.782pt}{0.800pt}}
\put(1144.0,126.0){\rule[-0.400pt]{0.800pt}{1.927pt}}
\put(1144.0,126.0){\rule[-0.400pt]{5.782pt}{0.800pt}}
\put(1168.0,121.0){\rule[-0.400pt]{0.800pt}{1.204pt}}
\put(1168.0,121.0){\rule[-0.400pt]{6.022pt}{0.800pt}}
\put(1193.0,118.0){\usebox{\plotpoint}}
\put(1193.0,118.0){\rule[-0.400pt]{5.782pt}{0.800pt}}
\put(1217.0,115.0){\usebox{\plotpoint}}
\put(1217.0,115.0){\rule[-0.400pt]{5.782pt}{0.800pt}}
\put(1241.0,114.0){\usebox{\plotpoint}}
\put(1241.0,114.0){\rule[-0.400pt]{6.022pt}{0.800pt}}
\put(1266.0,113.0){\usebox{\plotpoint}}
\put(1266.0,113.0){\rule[-0.400pt]{40.953pt}{0.800pt}}
\end{picture}
\vskip 1 in
\begin{center}
\bf{Figure 2}
\end{center}
\vskip .1 in
\noindent
Distribution of $E{\!\!\!\!/_T}$~~ for the process
$e^-\gamma\longrightarrow e^-e^+e^- + E{\!\!\!\!/_T}$~~, for 
two values of the SUSY breaking scale. We have set $\mu$ = 300, 
tan$\beta$ = 2, and $M/\Lambda$ = 2. All cuts except the 
$E{\!\!\!\!/_T}$~cut (set to zero) are as discussed in the text.
\newpage
\setlength{\unitlength}{0.240900pt}
\ifx\plotpoint\undefined\newsavebox{\plotpoint}\fi
\sbox{\plotpoint}{\rule[-0.200pt]{0.400pt}{0.400pt}}%
\begin{picture}(1500,900)(0,0)
\font\gnuplot=cmr10 at 10pt
\gnuplot
\sbox{\plotpoint}{\rule[-0.200pt]{0.400pt}{0.400pt}}%
\put(220.0,113.0){\rule[-0.200pt]{292.934pt}{0.400pt}}
\put(220.0,113.0){\rule[-0.200pt]{0.400pt}{184.048pt}}
\put(220.0,113.0){\rule[-0.200pt]{4.818pt}{0.400pt}}
\put(198,113){\makebox(0,0)[r]{0}}
\put(1416.0,113.0){\rule[-0.200pt]{4.818pt}{0.400pt}}
\put(220.0,240.0){\rule[-0.200pt]{4.818pt}{0.400pt}}
\put(198,240){\makebox(0,0)[r]{0.001}}
\put(1416.0,240.0){\rule[-0.200pt]{4.818pt}{0.400pt}}
\put(220.0,368.0){\rule[-0.200pt]{4.818pt}{0.400pt}}
\put(198,368){\makebox(0,0)[r]{0.002}}
\put(1416.0,368.0){\rule[-0.200pt]{4.818pt}{0.400pt}}
\put(220.0,495.0){\rule[-0.200pt]{4.818pt}{0.400pt}}
\put(198,495){\makebox(0,0)[r]{0.003}}
\put(1416.0,495.0){\rule[-0.200pt]{4.818pt}{0.400pt}}
\put(220.0,622.0){\rule[-0.200pt]{4.818pt}{0.400pt}}
\put(198,622){\makebox(0,0)[r]{0.004}}
\put(1416.0,622.0){\rule[-0.200pt]{4.818pt}{0.400pt}}
\put(220.0,750.0){\rule[-0.200pt]{4.818pt}{0.400pt}}
\put(198,750){\makebox(0,0)[r]{0.005}}
\put(1416.0,750.0){\rule[-0.200pt]{4.818pt}{0.400pt}}
\put(220.0,877.0){\rule[-0.200pt]{4.818pt}{0.400pt}}
\put(198,877){\makebox(0,0)[r]{0.006}}
\put(1416.0,877.0){\rule[-0.200pt]{4.818pt}{0.400pt}}
\put(220.0,113.0){\rule[-0.200pt]{0.400pt}{4.818pt}}
\put(220,68){\makebox(0,0){0}}
\put(220.0,857.0){\rule[-0.200pt]{0.400pt}{4.818pt}}
\put(463.0,113.0){\rule[-0.200pt]{0.400pt}{4.818pt}}
\put(463,68){\makebox(0,0){50}}
\put(463.0,857.0){\rule[-0.200pt]{0.400pt}{4.818pt}}
\put(706.0,113.0){\rule[-0.200pt]{0.400pt}{4.818pt}}
\put(706,68){\makebox(0,0){100}}
\put(706.0,857.0){\rule[-0.200pt]{0.400pt}{4.818pt}}
\put(950.0,113.0){\rule[-0.200pt]{0.400pt}{4.818pt}}
\put(950,68){\makebox(0,0){150}}
\put(950.0,857.0){\rule[-0.200pt]{0.400pt}{4.818pt}}
\put(1193.0,113.0){\rule[-0.200pt]{0.400pt}{4.818pt}}
\put(1193,68){\makebox(0,0){200}}
\put(1193.0,857.0){\rule[-0.200pt]{0.400pt}{4.818pt}}
\put(1436.0,113.0){\rule[-0.200pt]{0.400pt}{4.818pt}}
\put(1436,68){\makebox(0,0){250}}
\put(1436.0,857.0){\rule[-0.200pt]{0.400pt}{4.818pt}}
\put(220.0,113.0){\rule[-0.200pt]{292.934pt}{0.400pt}}
\put(1436.0,113.0){\rule[-0.200pt]{0.400pt}{184.048pt}}
\put(220.0,877.0){\rule[-0.200pt]{292.934pt}{0.400pt}}
\put(5,495){\makebox(0,0){$d\sigma$ (pb)}}
\put(828,23){\makebox(0,0){$E$ (GeV)}}
\put(220.0,113.0){\rule[-0.200pt]{0.400pt}{184.048pt}}
\put(1306,812){\makebox(0,0)[r]{1st lepton}}
\put(606,812){\makebox(0,0)[r]{$M$ = 50 TeV}}
\put(1328.0,812.0){\rule[-0.200pt]{15.899pt}{0.400pt}}
\put(220,308){\usebox{\plotpoint}}
\put(220.0,308.0){\rule[-0.200pt]{5.782pt}{0.400pt}}
\put(244.0,308.0){\rule[-0.200pt]{0.400pt}{113.223pt}}
\put(244.0,778.0){\rule[-0.200pt]{6.022pt}{0.400pt}}
\put(269.0,765.0){\rule[-0.200pt]{0.400pt}{3.132pt}}
\put(269.0,765.0){\rule[-0.200pt]{5.782pt}{0.400pt}}
\put(293.0,738.0){\rule[-0.200pt]{0.400pt}{6.504pt}}
\put(293.0,738.0){\rule[-0.200pt]{5.782pt}{0.400pt}}
\put(317.0,707.0){\rule[-0.200pt]{0.400pt}{7.468pt}}
\put(317.0,707.0){\rule[-0.200pt]{6.022pt}{0.400pt}}
\put(342.0,678.0){\rule[-0.200pt]{0.400pt}{6.986pt}}
\put(342.0,678.0){\rule[-0.200pt]{5.782pt}{0.400pt}}
\put(366.0,651.0){\rule[-0.200pt]{0.400pt}{6.504pt}}
\put(366.0,651.0){\rule[-0.200pt]{5.782pt}{0.400pt}}
\put(390.0,625.0){\rule[-0.200pt]{0.400pt}{6.263pt}}
\put(390.0,625.0){\rule[-0.200pt]{6.022pt}{0.400pt}}
\put(415.0,597.0){\rule[-0.200pt]{0.400pt}{6.745pt}}
\put(415.0,597.0){\rule[-0.200pt]{5.782pt}{0.400pt}}
\put(439.0,557.0){\rule[-0.200pt]{0.400pt}{9.636pt}}
\put(439.0,557.0){\rule[-0.200pt]{5.782pt}{0.400pt}}
\put(463.0,509.0){\rule[-0.200pt]{0.400pt}{11.563pt}}
\put(463.0,509.0){\rule[-0.200pt]{6.022pt}{0.400pt}}
\put(488.0,450.0){\rule[-0.200pt]{0.400pt}{14.213pt}}
\put(488.0,450.0){\rule[-0.200pt]{5.782pt}{0.400pt}}
\put(512.0,369.0){\rule[-0.200pt]{0.400pt}{19.513pt}}
\put(512.0,369.0){\rule[-0.200pt]{5.782pt}{0.400pt}}
\put(536.0,255.0){\rule[-0.200pt]{0.400pt}{27.463pt}}
\put(536.0,255.0){\rule[-0.200pt]{5.782pt}{0.400pt}}
\put(560.0,128.0){\rule[-0.200pt]{0.400pt}{30.594pt}}
\put(560.0,128.0){\rule[-0.200pt]{6.022pt}{0.400pt}}
\put(585.0,113.0){\rule[-0.200pt]{0.400pt}{3.613pt}}
\put(585.0,113.0){\rule[-0.200pt]{205.006pt}{0.400pt}}
\sbox{\plotpoint}{\rule[-0.400pt]{0.800pt}{0.800pt}}%
\put(1306,767){\makebox(0,0)[r]{2nd lepton}}
\put(1328.0,767.0){\rule[-0.400pt]{15.899pt}{0.800pt}}
\put(220,113){\usebox{\plotpoint}}
\put(220.0,113.0){\rule[-0.400pt]{5.782pt}{0.800pt}}
\put(244.0,113.0){\rule[-0.400pt]{0.800pt}{21.199pt}}
\put(244.0,201.0){\rule[-0.400pt]{6.022pt}{0.800pt}}
\put(269.0,201.0){\rule[-0.400pt]{0.800pt}{59.261pt}}
\put(269.0,447.0){\rule[-0.400pt]{5.782pt}{0.800pt}}
\put(293.0,447.0){\rule[-0.400pt]{0.800pt}{5.782pt}}
\put(293.0,471.0){\rule[-0.400pt]{5.782pt}{0.800pt}}
\put(317.0,443.0){\rule[-0.400pt]{0.800pt}{6.745pt}}
\put(317.0,443.0){\rule[-0.400pt]{6.022pt}{0.800pt}}
\put(342.0,429.0){\rule[-0.400pt]{0.800pt}{3.373pt}}
\put(342.0,429.0){\rule[-0.400pt]{5.782pt}{0.800pt}}
\put(366.0,406.0){\rule[-0.400pt]{0.800pt}{5.541pt}}
\put(366.0,406.0){\rule[-0.400pt]{5.782pt}{0.800pt}}
\put(390.0,390.0){\rule[-0.400pt]{0.800pt}{3.854pt}}
\put(390.0,390.0){\rule[-0.400pt]{6.022pt}{0.800pt}}
\put(415.0,375.0){\rule[-0.400pt]{0.800pt}{3.613pt}}
\put(415.0,375.0){\rule[-0.400pt]{5.782pt}{0.800pt}}
\put(439.0,364.0){\rule[-0.400pt]{0.800pt}{2.650pt}}
\put(439.0,364.0){\rule[-0.400pt]{5.782pt}{0.800pt}}
\put(463.0,354.0){\rule[-0.400pt]{0.800pt}{2.409pt}}
\put(463.0,354.0){\rule[-0.400pt]{6.022pt}{0.800pt}}
\put(488.0,342.0){\rule[-0.400pt]{0.800pt}{2.891pt}}
\put(488.0,342.0){\rule[-0.400pt]{5.782pt}{0.800pt}}
\put(512.0,328.0){\rule[-0.400pt]{0.800pt}{3.373pt}}
\put(512.0,328.0){\rule[-0.400pt]{5.782pt}{0.800pt}}
\put(536.0,321.0){\rule[-0.400pt]{0.800pt}{1.686pt}}
\put(536.0,321.0){\rule[-0.400pt]{5.782pt}{0.800pt}}
\put(560.0,316.0){\rule[-0.400pt]{0.800pt}{1.204pt}}
\put(560.0,316.0){\rule[-0.400pt]{6.022pt}{0.800pt}}
\put(585.0,306.0){\rule[-0.400pt]{0.800pt}{2.409pt}}
\put(585.0,306.0){\rule[-0.400pt]{5.782pt}{0.800pt}}
\put(609.0,301.0){\rule[-0.400pt]{0.800pt}{1.204pt}}
\put(609.0,301.0){\rule[-0.400pt]{5.782pt}{0.800pt}}
\put(633.0,295.0){\rule[-0.400pt]{0.800pt}{1.445pt}}
\put(633.0,295.0){\rule[-0.400pt]{6.022pt}{0.800pt}}
\put(658.0,288.0){\rule[-0.400pt]{0.800pt}{1.686pt}}
\put(658.0,288.0){\rule[-0.400pt]{5.782pt}{0.800pt}}
\put(682.0,282.0){\rule[-0.400pt]{0.800pt}{1.445pt}}
\put(682.0,282.0){\rule[-0.400pt]{5.782pt}{0.800pt}}
\put(706.0,275.0){\rule[-0.400pt]{0.800pt}{1.686pt}}
\put(706.0,275.0){\rule[-0.400pt]{6.022pt}{0.800pt}}
\put(731.0,270.0){\rule[-0.400pt]{0.800pt}{1.204pt}}
\put(731.0,270.0){\rule[-0.400pt]{5.782pt}{0.800pt}}
\put(755.0,260.0){\rule[-0.400pt]{0.800pt}{2.409pt}}
\put(755.0,260.0){\rule[-0.400pt]{5.782pt}{0.800pt}}
\put(779.0,254.0){\rule[-0.400pt]{0.800pt}{1.445pt}}
\put(779.0,254.0){\rule[-0.400pt]{6.022pt}{0.800pt}}
\put(804.0,245.0){\rule[-0.400pt]{0.800pt}{2.168pt}}
\put(804.0,245.0){\rule[-0.400pt]{5.782pt}{0.800pt}}
\put(828.0,239.0){\rule[-0.400pt]{0.800pt}{1.445pt}}
\put(828.0,239.0){\rule[-0.400pt]{5.782pt}{0.800pt}}
\put(852.0,231.0){\rule[-0.400pt]{0.800pt}{1.927pt}}
\put(852.0,231.0){\rule[-0.400pt]{6.022pt}{0.800pt}}
\put(877.0,223.0){\rule[-0.400pt]{0.800pt}{1.927pt}}
\put(877.0,223.0){\rule[-0.400pt]{5.782pt}{0.800pt}}
\put(901.0,211.0){\rule[-0.400pt]{0.800pt}{2.891pt}}
\put(901.0,211.0){\rule[-0.400pt]{5.782pt}{0.800pt}}
\put(925.0,207.0){\rule[-0.400pt]{0.800pt}{0.964pt}}
\put(925.0,207.0){\rule[-0.400pt]{6.022pt}{0.800pt}}
\put(950.0,200.0){\rule[-0.400pt]{0.800pt}{1.686pt}}
\put(950.0,200.0){\rule[-0.400pt]{5.782pt}{0.800pt}}
\put(974.0,190.0){\rule[-0.400pt]{0.800pt}{2.409pt}}
\put(974.0,190.0){\rule[-0.400pt]{5.782pt}{0.800pt}}
\put(998.0,187.0){\usebox{\plotpoint}}
\put(998.0,187.0){\rule[-0.400pt]{6.022pt}{0.800pt}}
\put(1023.0,178.0){\rule[-0.400pt]{0.800pt}{2.168pt}}
\put(1023.0,178.0){\rule[-0.400pt]{5.782pt}{0.800pt}}
\put(1047.0,174.0){\rule[-0.400pt]{0.800pt}{0.964pt}}
\put(1047.0,174.0){\rule[-0.400pt]{5.782pt}{0.800pt}}
\put(1071.0,166.0){\rule[-0.400pt]{0.800pt}{1.927pt}}
\put(1071.0,166.0){\rule[-0.400pt]{6.022pt}{0.800pt}}
\put(1096.0,162.0){\rule[-0.400pt]{0.800pt}{0.964pt}}
\put(1096.0,162.0){\rule[-0.400pt]{5.782pt}{0.800pt}}
\put(1120.0,153.0){\rule[-0.400pt]{0.800pt}{2.168pt}}
\put(1120.0,153.0){\rule[-0.400pt]{5.782pt}{0.800pt}}
\put(1144.0,145.0){\rule[-0.400pt]{0.800pt}{1.927pt}}
\put(1144.0,145.0){\rule[-0.400pt]{5.782pt}{0.800pt}}
\put(1168.0,140.0){\rule[-0.400pt]{0.800pt}{1.204pt}}
\put(1168.0,140.0){\rule[-0.400pt]{6.022pt}{0.800pt}}
\put(1193.0,132.0){\rule[-0.400pt]{0.800pt}{1.927pt}}
\put(1193.0,132.0){\rule[-0.400pt]{5.782pt}{0.800pt}}
\put(1217.0,128.0){\rule[-0.400pt]{0.800pt}{0.964pt}}
\put(1217.0,128.0){\rule[-0.400pt]{5.782pt}{0.800pt}}
\put(1241.0,123.0){\rule[-0.400pt]{0.800pt}{1.204pt}}
\put(1241.0,123.0){\rule[-0.400pt]{6.022pt}{0.800pt}}
\put(1266.0,120.0){\usebox{\plotpoint}}
\put(1266.0,120.0){\rule[-0.400pt]{5.782pt}{0.800pt}}
\put(1290.0,118.0){\usebox{\plotpoint}}
\put(1290.0,118.0){\rule[-0.400pt]{5.782pt}{0.800pt}}
\put(1314.0,115.0){\usebox{\plotpoint}}
\put(1314.0,115.0){\rule[-0.400pt]{6.022pt}{0.800pt}}
\put(1339.0,114.0){\usebox{\plotpoint}}
\put(1339.0,114.0){\rule[-0.400pt]{5.782pt}{0.800pt}}
\put(1363.0,113.0){\usebox{\plotpoint}}
\put(1363.0,113.0){\rule[-0.400pt]{17.586pt}{0.800pt}}
\sbox{\plotpoint}{\rule[-0.600pt]{1.200pt}{1.200pt}}%
\put(1306,722){\makebox(0,0)[r]{3rd lepton}}
\put(1328.0,722.0){\rule[-0.600pt]{15.899pt}{1.200pt}}
\put(220,113){\usebox{\plotpoint}}
\put(220.0,113.0){\rule[-0.600pt]{5.782pt}{1.200pt}}
\put(244.0,113.0){\rule[-0.600pt]{1.200pt}{11.563pt}}
\put(244.0,161.0){\rule[-0.600pt]{6.022pt}{1.200pt}}
\put(269.0,161.0){\rule[-0.600pt]{1.200pt}{35.171pt}}
\put(269.0,307.0){\rule[-0.600pt]{5.782pt}{1.200pt}}
\put(293.0,307.0){\rule[-0.600pt]{1.200pt}{7.950pt}}
\put(293.0,340.0){\rule[-0.600pt]{5.782pt}{1.200pt}}
\put(317.0,340.0){\usebox{\plotpoint}}
\put(317.0,343.0){\rule[-0.600pt]{6.022pt}{1.200pt}}
\put(342.0,343.0){\usebox{\plotpoint}}
\put(342.0,344.0){\rule[-0.600pt]{5.782pt}{1.200pt}}
\put(366.0,338.0){\rule[-0.600pt]{1.200pt}{1.445pt}}
\put(366.0,338.0){\rule[-0.600pt]{5.782pt}{1.200pt}}
\put(390.0,334.0){\usebox{\plotpoint}}
\put(390.0,334.0){\rule[-0.600pt]{6.022pt}{1.200pt}}
\put(415.0,333.0){\usebox{\plotpoint}}
\put(415.0,333.0){\rule[-0.600pt]{11.563pt}{1.200pt}}
\put(463.0,331.0){\usebox{\plotpoint}}
\put(463.0,331.0){\rule[-0.600pt]{6.022pt}{1.200pt}}
\put(488.0,326.0){\rule[-0.600pt]{1.200pt}{1.204pt}}
\put(488.0,326.0){\rule[-0.600pt]{5.782pt}{1.200pt}}
\put(512.0,325.0){\usebox{\plotpoint}}
\put(512.0,325.0){\rule[-0.600pt]{5.782pt}{1.200pt}}
\put(536.0,321.0){\usebox{\plotpoint}}
\put(536.0,321.0){\rule[-0.600pt]{11.804pt}{1.200pt}}
\put(585.0,317.0){\usebox{\plotpoint}}
\put(585.0,317.0){\rule[-0.600pt]{5.782pt}{1.200pt}}
\put(609.0,316.0){\usebox{\plotpoint}}
\put(609.0,316.0){\rule[-0.600pt]{5.782pt}{1.200pt}}
\put(633.0,309.0){\rule[-0.600pt]{1.200pt}{1.686pt}}
\put(633.0,309.0){\rule[-0.600pt]{6.022pt}{1.200pt}}
\put(658.0,307.0){\usebox{\plotpoint}}
\put(658.0,307.0){\rule[-0.600pt]{5.782pt}{1.200pt}}
\put(682.0,303.0){\usebox{\plotpoint}}
\put(682.0,303.0){\rule[-0.600pt]{5.782pt}{1.200pt}}
\put(706.0,302.0){\usebox{\plotpoint}}
\put(706.0,302.0){\rule[-0.600pt]{6.022pt}{1.200pt}}
\put(731.0,297.0){\rule[-0.600pt]{1.200pt}{1.204pt}}
\put(731.0,297.0){\rule[-0.600pt]{5.782pt}{1.200pt}}
\put(755.0,294.0){\usebox{\plotpoint}}
\put(755.0,294.0){\rule[-0.600pt]{5.782pt}{1.200pt}}
\put(779.0,285.0){\rule[-0.600pt]{1.200pt}{2.168pt}}
\put(779.0,285.0){\rule[-0.600pt]{6.022pt}{1.200pt}}
\put(804.0,283.0){\usebox{\plotpoint}}
\put(804.0,283.0){\rule[-0.600pt]{5.782pt}{1.200pt}}
\put(828.0,275.0){\rule[-0.600pt]{1.200pt}{1.927pt}}
\put(828.0,275.0){\rule[-0.600pt]{5.782pt}{1.200pt}}
\put(852.0,273.0){\usebox{\plotpoint}}
\put(852.0,273.0){\rule[-0.600pt]{6.022pt}{1.200pt}}
\put(877.0,265.0){\rule[-0.600pt]{1.200pt}{1.927pt}}
\put(877.0,265.0){\rule[-0.600pt]{5.782pt}{1.200pt}}
\put(901.0,261.0){\usebox{\plotpoint}}
\put(901.0,261.0){\rule[-0.600pt]{5.782pt}{1.200pt}}
\put(925.0,249.0){\rule[-0.600pt]{1.200pt}{2.891pt}}
\put(925.0,249.0){\rule[-0.600pt]{6.022pt}{1.200pt}}
\put(950.0,241.0){\rule[-0.600pt]{1.200pt}{1.927pt}}
\put(950.0,241.0){\rule[-0.600pt]{5.782pt}{1.200pt}}
\put(974.0,233.0){\rule[-0.600pt]{1.200pt}{1.927pt}}
\put(974.0,233.0){\rule[-0.600pt]{5.782pt}{1.200pt}}
\put(998.0,223.0){\rule[-0.600pt]{1.200pt}{2.409pt}}
\put(998.0,223.0){\rule[-0.600pt]{6.022pt}{1.200pt}}
\put(1023.0,213.0){\rule[-0.600pt]{1.200pt}{2.409pt}}
\put(1023.0,213.0){\rule[-0.600pt]{5.782pt}{1.200pt}}
\put(1047.0,197.0){\rule[-0.600pt]{1.200pt}{3.854pt}}
\put(1047.0,197.0){\rule[-0.600pt]{5.782pt}{1.200pt}}
\put(1071.0,187.0){\rule[-0.600pt]{1.200pt}{2.409pt}}
\put(1071.0,187.0){\rule[-0.600pt]{6.022pt}{1.200pt}}
\put(1096.0,172.0){\rule[-0.600pt]{1.200pt}{3.613pt}}
\put(1096.0,172.0){\rule[-0.600pt]{5.782pt}{1.200pt}}
\put(1120.0,167.0){\rule[-0.600pt]{1.200pt}{1.204pt}}
\put(1120.0,167.0){\rule[-0.600pt]{5.782pt}{1.200pt}}
\put(1144.0,160.0){\rule[-0.600pt]{1.200pt}{1.686pt}}
\put(1144.0,160.0){\rule[-0.600pt]{5.782pt}{1.200pt}}
\put(1168.0,148.0){\rule[-0.600pt]{1.200pt}{2.891pt}}
\put(1168.0,148.0){\rule[-0.600pt]{6.022pt}{1.200pt}}
\put(1193.0,143.0){\rule[-0.600pt]{1.200pt}{1.204pt}}
\put(1193.0,143.0){\rule[-0.600pt]{5.782pt}{1.200pt}}
\put(1217.0,135.0){\rule[-0.600pt]{1.200pt}{1.927pt}}
\put(1217.0,135.0){\rule[-0.600pt]{5.782pt}{1.200pt}}
\put(1241.0,129.0){\rule[-0.600pt]{1.200pt}{1.445pt}}
\put(1241.0,129.0){\rule[-0.600pt]{6.022pt}{1.200pt}}
\put(1266.0,125.0){\usebox{\plotpoint}}
\put(1266.0,125.0){\rule[-0.600pt]{5.782pt}{1.200pt}}
\put(1290.0,120.0){\rule[-0.600pt]{1.200pt}{1.204pt}}
\put(1290.0,120.0){\rule[-0.600pt]{5.782pt}{1.200pt}}
\put(1314.0,117.0){\usebox{\plotpoint}}
\put(1314.0,117.0){\rule[-0.600pt]{6.022pt}{1.200pt}}
\put(1339.0,114.0){\usebox{\plotpoint}}
\put(1339.0,114.0){\rule[-0.600pt]{5.782pt}{1.200pt}}
\put(1363.0,113.0){\usebox{\plotpoint}}
\put(1363.0,113.0){\rule[-0.600pt]{17.586pt}{1.200pt}}
\end{picture}
\begin{center}
\bf{Figure 3(a)}
\end{center}
\setlength{\unitlength}{0.240900pt}
\ifx\plotpoint\undefined\newsavebox{\plotpoint}\fi
\sbox{\plotpoint}{\rule[-0.200pt]{0.400pt}{0.400pt}}%
\begin{picture}(1500,900)(0,0)
\font\gnuplot=cmr10 at 10pt
\gnuplot
\sbox{\plotpoint}{\rule[-0.200pt]{0.400pt}{0.400pt}}%
\put(220.0,113.0){\rule[-0.200pt]{292.934pt}{0.400pt}}
\put(220.0,113.0){\rule[-0.200pt]{0.400pt}{184.048pt}}
\put(220.0,113.0){\rule[-0.200pt]{4.818pt}{0.400pt}}
\put(198,113){\makebox(0,0)[r]{0}}
\put(1416.0,113.0){\rule[-0.200pt]{4.818pt}{0.400pt}}
\put(220.0,209.0){\rule[-0.200pt]{4.818pt}{0.400pt}}
\put(198,209){\makebox(0,0)[r]{0.0005}}
\put(1416.0,209.0){\rule[-0.200pt]{4.818pt}{0.400pt}}
\put(220.0,304.0){\rule[-0.200pt]{4.818pt}{0.400pt}}
\put(198,304){\makebox(0,0)[r]{0.001}}
\put(1416.0,304.0){\rule[-0.200pt]{4.818pt}{0.400pt}}
\put(220.0,400.0){\rule[-0.200pt]{4.818pt}{0.400pt}}
\put(198,400){\makebox(0,0)[r]{0.0015}}
\put(1416.0,400.0){\rule[-0.200pt]{4.818pt}{0.400pt}}
\put(220.0,495.0){\rule[-0.200pt]{4.818pt}{0.400pt}}
\put(198,495){\makebox(0,0)[r]{0.002}}
\put(1416.0,495.0){\rule[-0.200pt]{4.818pt}{0.400pt}}
\put(220.0,591.0){\rule[-0.200pt]{4.818pt}{0.400pt}}
\put(198,591){\makebox(0,0)[r]{0.0025}}
\put(1416.0,591.0){\rule[-0.200pt]{4.818pt}{0.400pt}}
\put(220.0,686.0){\rule[-0.200pt]{4.818pt}{0.400pt}}
\put(198,686){\makebox(0,0)[r]{0.003}}
\put(1416.0,686.0){\rule[-0.200pt]{4.818pt}{0.400pt}}
\put(220.0,782.0){\rule[-0.200pt]{4.818pt}{0.400pt}}
\put(198,782){\makebox(0,0)[r]{0.0035}}
\put(1416.0,782.0){\rule[-0.200pt]{4.818pt}{0.400pt}}
\put(220.0,877.0){\rule[-0.200pt]{4.818pt}{0.400pt}}
\put(198,877){\makebox(0,0)[r]{0.004}}
\put(1416.0,877.0){\rule[-0.200pt]{4.818pt}{0.400pt}}
\put(220.0,113.0){\rule[-0.200pt]{0.400pt}{4.818pt}}
\put(220,68){\makebox(0,0){0}}
\put(220.0,857.0){\rule[-0.200pt]{0.400pt}{4.818pt}}
\put(463.0,113.0){\rule[-0.200pt]{0.400pt}{4.818pt}}
\put(463,68){\makebox(0,0){50}}
\put(463.0,857.0){\rule[-0.200pt]{0.400pt}{4.818pt}}
\put(706.0,113.0){\rule[-0.200pt]{0.400pt}{4.818pt}}
\put(706,68){\makebox(0,0){100}}
\put(706.0,857.0){\rule[-0.200pt]{0.400pt}{4.818pt}}
\put(950.0,113.0){\rule[-0.200pt]{0.400pt}{4.818pt}}
\put(950,68){\makebox(0,0){150}}
\put(950.0,857.0){\rule[-0.200pt]{0.400pt}{4.818pt}}
\put(1193.0,113.0){\rule[-0.200pt]{0.400pt}{4.818pt}}
\put(1193,68){\makebox(0,0){200}}
\put(1193.0,857.0){\rule[-0.200pt]{0.400pt}{4.818pt}}
\put(1436.0,113.0){\rule[-0.200pt]{0.400pt}{4.818pt}}
\put(1436,68){\makebox(0,0){250}}
\put(1436.0,857.0){\rule[-0.200pt]{0.400pt}{4.818pt}}
\put(220.0,113.0){\rule[-0.200pt]{292.934pt}{0.400pt}}
\put(1436.0,113.0){\rule[-0.200pt]{0.400pt}{184.048pt}}
\put(220.0,877.0){\rule[-0.200pt]{292.934pt}{0.400pt}}
\put(1,495){\makebox(0,0){$d\sigma$ (pb)}}
\put(828,23){\makebox(0,0){$E$ (GeV)}}
\put(220.0,113.0){\rule[-0.200pt]{0.400pt}{184.048pt}}
\put(1306,812){\makebox(0,0)[r]{1st lepton}}
\put(906,812){\makebox(0,0)[r]{$M$ = 100 TeV}}
\put(1328.0,812.0){\rule[-0.200pt]{15.899pt}{0.400pt}}
\put(220,113){\usebox{\plotpoint}}
\put(220.0,113.0){\rule[-0.200pt]{17.586pt}{0.400pt}}
\put(293.0,113.0){\rule[-0.200pt]{0.400pt}{29.149pt}}
\put(293.0,234.0){\rule[-0.200pt]{5.782pt}{0.400pt}}
\put(317.0,234.0){\rule[-0.200pt]{0.400pt}{85.038pt}}
\put(317.0,587.0){\rule[-0.200pt]{6.022pt}{0.400pt}}
\put(342.0,587.0){\rule[-0.200pt]{0.400pt}{33.726pt}}
\put(342.0,727.0){\rule[-0.200pt]{5.782pt}{0.400pt}}
\put(366.0,727.0){\rule[-0.200pt]{0.400pt}{12.045pt}}
\put(366.0,777.0){\rule[-0.200pt]{5.782pt}{0.400pt}}
\put(390.0,777.0){\rule[-0.200pt]{0.400pt}{2.650pt}}
\put(390.0,788.0){\rule[-0.200pt]{6.022pt}{0.400pt}}
\put(415.0,761.0){\rule[-0.200pt]{0.400pt}{6.504pt}}
\put(415.0,761.0){\rule[-0.200pt]{5.782pt}{0.400pt}}
\put(439.0,729.0){\rule[-0.200pt]{0.400pt}{7.709pt}}
\put(439.0,729.0){\rule[-0.200pt]{5.782pt}{0.400pt}}
\put(463.0,673.0){\rule[-0.200pt]{0.400pt}{13.490pt}}
\put(463.0,673.0){\rule[-0.200pt]{6.022pt}{0.400pt}}
\put(488.0,612.0){\rule[-0.200pt]{0.400pt}{14.695pt}}
\put(488.0,612.0){\rule[-0.200pt]{5.782pt}{0.400pt}}
\put(512.0,530.0){\rule[-0.200pt]{0.400pt}{19.754pt}}
\put(512.0,530.0){\rule[-0.200pt]{5.782pt}{0.400pt}}
\put(536.0,437.0){\rule[-0.200pt]{0.400pt}{22.404pt}}
\put(536.0,437.0){\rule[-0.200pt]{5.782pt}{0.400pt}}
\put(560.0,320.0){\rule[-0.200pt]{0.400pt}{28.185pt}}
\put(560.0,320.0){\rule[-0.200pt]{6.022pt}{0.400pt}}
\put(585.0,201.0){\rule[-0.200pt]{0.400pt}{28.667pt}}
\put(585.0,201.0){\rule[-0.200pt]{5.782pt}{0.400pt}}
\put(609.0,120.0){\rule[-0.200pt]{0.400pt}{19.513pt}}
\put(609.0,120.0){\rule[-0.200pt]{5.782pt}{0.400pt}}
\put(633.0,113.0){\rule[-0.200pt]{0.400pt}{1.686pt}}
\put(633.0,113.0){\rule[-0.200pt]{193.443pt}{0.400pt}}
\sbox{\plotpoint}{\rule[-0.400pt]{0.800pt}{0.800pt}}%
\put(1306,767){\makebox(0,0)[r]{2nd lepton}}
\put(1328.0,767.0){\rule[-0.400pt]{15.899pt}{0.800pt}}
\put(220,113){\usebox{\plotpoint}}
\put(220.0,113.0){\rule[-0.400pt]{29.390pt}{0.800pt}}
\put(342.0,113.0){\usebox{\plotpoint}}
\put(342.0,114.0){\rule[-0.400pt]{5.782pt}{0.800pt}}
\put(366.0,114.0){\rule[-0.400pt]{0.800pt}{13.731pt}}
\put(366.0,171.0){\rule[-0.400pt]{5.782pt}{0.800pt}}
\put(390.0,171.0){\rule[-0.400pt]{0.800pt}{26.981pt}}
\put(390.0,283.0){\rule[-0.400pt]{6.022pt}{0.800pt}}
\put(415.0,283.0){\rule[-0.400pt]{0.800pt}{15.658pt}}
\put(415.0,348.0){\rule[-0.400pt]{5.782pt}{0.800pt}}
\put(439.0,348.0){\rule[-0.400pt]{0.800pt}{8.672pt}}
\put(439.0,384.0){\rule[-0.400pt]{5.782pt}{0.800pt}}
\put(463.0,384.0){\rule[-0.400pt]{0.800pt}{5.059pt}}
\put(463.0,405.0){\rule[-0.400pt]{6.022pt}{0.800pt}}
\put(488.0,405.0){\rule[-0.400pt]{0.800pt}{3.373pt}}
\put(488.0,419.0){\rule[-0.400pt]{5.782pt}{0.800pt}}
\put(512.0,419.0){\usebox{\plotpoint}}
\put(512.0,422.0){\rule[-0.400pt]{11.563pt}{0.800pt}}
\put(560.0,419.0){\usebox{\plotpoint}}
\put(560.0,419.0){\rule[-0.400pt]{6.022pt}{0.800pt}}
\put(585.0,411.0){\rule[-0.400pt]{0.800pt}{1.927pt}}
\put(585.0,411.0){\rule[-0.400pt]{5.782pt}{0.800pt}}
\put(609.0,406.0){\rule[-0.400pt]{0.800pt}{1.204pt}}
\put(609.0,406.0){\rule[-0.400pt]{5.782pt}{0.800pt}}
\put(633.0,394.0){\rule[-0.400pt]{0.800pt}{2.891pt}}
\put(633.0,394.0){\rule[-0.400pt]{6.022pt}{0.800pt}}
\put(658.0,385.0){\rule[-0.400pt]{0.800pt}{2.168pt}}
\put(658.0,385.0){\rule[-0.400pt]{5.782pt}{0.800pt}}
\put(682.0,371.0){\rule[-0.400pt]{0.800pt}{3.373pt}}
\put(682.0,371.0){\rule[-0.400pt]{5.782pt}{0.800pt}}
\put(706.0,359.0){\rule[-0.400pt]{0.800pt}{2.891pt}}
\put(706.0,359.0){\rule[-0.400pt]{6.022pt}{0.800pt}}
\put(731.0,349.0){\rule[-0.400pt]{0.800pt}{2.409pt}}
\put(731.0,349.0){\rule[-0.400pt]{5.782pt}{0.800pt}}
\put(755.0,333.0){\rule[-0.400pt]{0.800pt}{3.854pt}}
\put(755.0,333.0){\rule[-0.400pt]{5.782pt}{0.800pt}}
\put(779.0,321.0){\rule[-0.400pt]{0.800pt}{2.891pt}}
\put(779.0,321.0){\rule[-0.400pt]{6.022pt}{0.800pt}}
\put(804.0,309.0){\rule[-0.400pt]{0.800pt}{2.891pt}}
\put(804.0,309.0){\rule[-0.400pt]{5.782pt}{0.800pt}}
\put(828.0,292.0){\rule[-0.400pt]{0.800pt}{4.095pt}}
\put(828.0,292.0){\rule[-0.400pt]{5.782pt}{0.800pt}}
\put(852.0,277.0){\rule[-0.400pt]{0.800pt}{3.613pt}}
\put(852.0,277.0){\rule[-0.400pt]{6.022pt}{0.800pt}}
\put(877.0,265.0){\rule[-0.400pt]{0.800pt}{2.891pt}}
\put(877.0,265.0){\rule[-0.400pt]{5.782pt}{0.800pt}}
\put(901.0,248.0){\rule[-0.400pt]{0.800pt}{4.095pt}}
\put(901.0,248.0){\rule[-0.400pt]{5.782pt}{0.800pt}}
\put(925.0,233.0){\rule[-0.400pt]{0.800pt}{3.613pt}}
\put(925.0,233.0){\rule[-0.400pt]{6.022pt}{0.800pt}}
\put(950.0,219.0){\rule[-0.400pt]{0.800pt}{3.373pt}}
\put(950.0,219.0){\rule[-0.400pt]{5.782pt}{0.800pt}}
\put(974.0,202.0){\rule[-0.400pt]{0.800pt}{4.095pt}}
\put(974.0,202.0){\rule[-0.400pt]{5.782pt}{0.800pt}}
\put(998.0,181.0){\rule[-0.400pt]{0.800pt}{5.059pt}}
\put(998.0,181.0){\rule[-0.400pt]{6.022pt}{0.800pt}}
\put(1023.0,166.0){\rule[-0.400pt]{0.800pt}{3.613pt}}
\put(1023.0,166.0){\rule[-0.400pt]{5.782pt}{0.800pt}}
\put(1047.0,148.0){\rule[-0.400pt]{0.800pt}{4.336pt}}
\put(1047.0,148.0){\rule[-0.400pt]{5.782pt}{0.800pt}}
\put(1071.0,137.0){\rule[-0.400pt]{0.800pt}{2.650pt}}
\put(1071.0,137.0){\rule[-0.400pt]{6.022pt}{0.800pt}}
\put(1096.0,127.0){\rule[-0.400pt]{0.800pt}{2.409pt}}
\put(1096.0,127.0){\rule[-0.400pt]{5.782pt}{0.800pt}}
\put(1120.0,120.0){\rule[-0.400pt]{0.800pt}{1.686pt}}
\put(1120.0,120.0){\rule[-0.400pt]{5.782pt}{0.800pt}}
\put(1144.0,115.0){\rule[-0.400pt]{0.800pt}{1.204pt}}
\put(1144.0,115.0){\rule[-0.400pt]{5.782pt}{0.800pt}}
\put(1168.0,113.0){\usebox{\plotpoint}}
\put(1168.0,113.0){\rule[-0.400pt]{64.561pt}{0.800pt}}
\sbox{\plotpoint}{\rule[-0.600pt]{1.200pt}{1.200pt}}%
\put(1306,722){\makebox(0,0)[r]{3rd lepton}}
\put(1328.0,722.0){\rule[-0.600pt]{15.899pt}{1.200pt}}
\put(220,113){\usebox{\plotpoint}}
\put(220.0,113.0){\rule[-0.600pt]{29.390pt}{1.200pt}}
\put(342.0,113.0){\usebox{\plotpoint}}
\put(342.0,117.0){\rule[-0.600pt]{5.782pt}{1.200pt}}
\put(366.0,117.0){\rule[-0.600pt]{1.200pt}{11.322pt}}
\put(366.0,164.0){\rule[-0.600pt]{5.782pt}{1.200pt}}
\put(390.0,164.0){\rule[-0.600pt]{1.200pt}{17.345pt}}
\put(390.0,236.0){\rule[-0.600pt]{6.022pt}{1.200pt}}
\put(415.0,236.0){\rule[-0.600pt]{1.200pt}{15.177pt}}
\put(415.0,299.0){\rule[-0.600pt]{5.782pt}{1.200pt}}
\put(439.0,299.0){\rule[-0.600pt]{1.200pt}{12.768pt}}
\put(439.0,352.0){\rule[-0.600pt]{5.782pt}{1.200pt}}
\put(463.0,352.0){\rule[-0.600pt]{1.200pt}{7.468pt}}
\put(463.0,383.0){\rule[-0.600pt]{6.022pt}{1.200pt}}
\put(488.0,383.0){\rule[-0.600pt]{1.200pt}{5.541pt}}
\put(488.0,406.0){\rule[-0.600pt]{5.782pt}{1.200pt}}
\put(512.0,406.0){\rule[-0.600pt]{1.200pt}{3.854pt}}
\put(512.0,422.0){\rule[-0.600pt]{5.782pt}{1.200pt}}
\put(536.0,422.0){\rule[-0.600pt]{1.200pt}{1.204pt}}
\put(536.0,427.0){\rule[-0.600pt]{5.782pt}{1.200pt}}
\put(560.0,427.0){\usebox{\plotpoint}}
\put(560.0,429.0){\rule[-0.600pt]{11.804pt}{1.200pt}}
\put(609.0,426.0){\usebox{\plotpoint}}
\put(609.0,426.0){\rule[-0.600pt]{5.782pt}{1.200pt}}
\put(633.0,418.0){\rule[-0.600pt]{1.200pt}{1.927pt}}
\put(633.0,418.0){\rule[-0.600pt]{6.022pt}{1.200pt}}
\put(658.0,411.0){\rule[-0.600pt]{1.200pt}{1.686pt}}
\put(658.0,411.0){\rule[-0.600pt]{5.782pt}{1.200pt}}
\put(682.0,401.0){\rule[-0.600pt]{1.200pt}{2.409pt}}
\put(682.0,401.0){\rule[-0.600pt]{5.782pt}{1.200pt}}
\put(706.0,384.0){\rule[-0.600pt]{1.200pt}{4.095pt}}
\put(706.0,384.0){\rule[-0.600pt]{6.022pt}{1.200pt}}
\put(731.0,374.0){\rule[-0.600pt]{1.200pt}{2.409pt}}
\put(731.0,374.0){\rule[-0.600pt]{5.782pt}{1.200pt}}
\put(755.0,353.0){\rule[-0.600pt]{1.200pt}{5.059pt}}
\put(755.0,353.0){\rule[-0.600pt]{5.782pt}{1.200pt}}
\put(779.0,339.0){\rule[-0.600pt]{1.200pt}{3.373pt}}
\put(779.0,339.0){\rule[-0.600pt]{6.022pt}{1.200pt}}
\put(804.0,320.0){\rule[-0.600pt]{1.200pt}{4.577pt}}
\put(804.0,320.0){\rule[-0.600pt]{5.782pt}{1.200pt}}
\put(828.0,295.0){\rule[-0.600pt]{1.200pt}{6.022pt}}
\put(828.0,295.0){\rule[-0.600pt]{5.782pt}{1.200pt}}
\put(852.0,275.0){\rule[-0.600pt]{1.200pt}{4.818pt}}
\put(852.0,275.0){\rule[-0.600pt]{6.022pt}{1.200pt}}
\put(877.0,254.0){\rule[-0.600pt]{1.200pt}{5.059pt}}
\put(877.0,254.0){\rule[-0.600pt]{5.782pt}{1.200pt}}
\put(901.0,235.0){\rule[-0.600pt]{1.200pt}{4.577pt}}
\put(901.0,235.0){\rule[-0.600pt]{5.782pt}{1.200pt}}
\put(925.0,216.0){\rule[-0.600pt]{1.200pt}{4.577pt}}
\put(925.0,216.0){\rule[-0.600pt]{6.022pt}{1.200pt}}
\put(950.0,202.0){\rule[-0.600pt]{1.200pt}{3.373pt}}
\put(950.0,202.0){\rule[-0.600pt]{5.782pt}{1.200pt}}
\put(974.0,185.0){\rule[-0.600pt]{1.200pt}{4.095pt}}
\put(974.0,185.0){\rule[-0.600pt]{5.782pt}{1.200pt}}
\put(998.0,172.0){\rule[-0.600pt]{1.200pt}{3.132pt}}
\put(998.0,172.0){\rule[-0.600pt]{6.022pt}{1.200pt}}
\put(1023.0,161.0){\rule[-0.600pt]{1.200pt}{2.650pt}}
\put(1023.0,161.0){\rule[-0.600pt]{5.782pt}{1.200pt}}
\put(1047.0,149.0){\rule[-0.600pt]{1.200pt}{2.891pt}}
\put(1047.0,149.0){\rule[-0.600pt]{5.782pt}{1.200pt}}
\put(1071.0,140.0){\rule[-0.600pt]{1.200pt}{2.168pt}}
\put(1071.0,140.0){\rule[-0.600pt]{6.022pt}{1.200pt}}
\put(1096.0,132.0){\rule[-0.600pt]{1.200pt}{1.927pt}}
\put(1096.0,132.0){\rule[-0.600pt]{5.782pt}{1.200pt}}
\put(1120.0,125.0){\rule[-0.600pt]{1.200pt}{1.686pt}}
\put(1120.0,125.0){\rule[-0.600pt]{5.782pt}{1.200pt}}
\put(1144.0,121.0){\usebox{\plotpoint}}
\put(1144.0,121.0){\rule[-0.600pt]{5.782pt}{1.200pt}}
\put(1168.0,117.0){\usebox{\plotpoint}}
\put(1168.0,117.0){\rule[-0.600pt]{6.022pt}{1.200pt}}
\put(1193.0,115.0){\usebox{\plotpoint}}
\put(1193.0,115.0){\rule[-0.600pt]{5.782pt}{1.200pt}}
\put(1217.0,113.0){\usebox{\plotpoint}}
\put(1217.0,113.0){\rule[-0.600pt]{52.757pt}{1.200pt}}
\end{picture}
\begin{center}
\bf{Figure 3(b)}
\end{center}
\vskip 1 in
Both the figures represent the distribution of the 
energies of three leptons for the process 
$e^-\gamma\longrightarrow e^-e^+e^- + E{\!\!\!\!/_T}$~~, for two 
values of the SUSY breaking scale (a) $M$ = 50 TeV (b) $M$ = 100 TeV. 
We have set $\mu$= 300, 
tan$\beta$ = 2, $M/\Lambda$ = 2. All cuts except the softness cut 
are as discussed in the text; the softness cut has been put to 
zero. The lepton number assignment follows the convention: 
$e^-\gamma\longrightarrow {\tilde e_R}(1)\chi_1^0$,
$\chi_1^0\longrightarrow {\tilde e_R}(2)e(1)$, 
${\tilde e_R}(1)\longrightarrow e(2)G$,
${\tilde e_R}(2)\longrightarrow e(3)G$.  
\newpage
\setlength{\unitlength}{0.240900pt}
\ifx\plotpoint\undefined\newsavebox{\plotpoint}\fi
\sbox{\plotpoint}{\rule[-0.200pt]{0.400pt}{0.400pt}}%
\begin{picture}(1500,900)(0,0)
\font\gnuplot=cmr10 at 10pt
\gnuplot
\sbox{\plotpoint}{\rule[-0.200pt]{0.400pt}{0.400pt}}%
\put(220.0,113.0){\rule[-0.200pt]{0.400pt}{184.048pt}}
\put(220.0,113.0){\rule[-0.200pt]{4.818pt}{0.400pt}}
\put(198,113){\makebox(0,0)[r]{0.0006}}
\put(1416.0,113.0){\rule[-0.200pt]{4.818pt}{0.400pt}}
\put(220.0,209.0){\rule[-0.200pt]{4.818pt}{0.400pt}}
\put(198,209){\makebox(0,0)[r]{0.0007}}
\put(1416.0,209.0){\rule[-0.200pt]{4.818pt}{0.400pt}}
\put(220.0,304.0){\rule[-0.200pt]{4.818pt}{0.400pt}}
\put(198,304){\makebox(0,0)[r]{0.0008}}
\put(1416.0,304.0){\rule[-0.200pt]{4.818pt}{0.400pt}}
\put(220.0,400.0){\rule[-0.200pt]{4.818pt}{0.400pt}}
\put(198,400){\makebox(0,0)[r]{0.0009}}
\put(1416.0,400.0){\rule[-0.200pt]{4.818pt}{0.400pt}}
\put(220.0,495.0){\rule[-0.200pt]{4.818pt}{0.400pt}}
\put(198,495){\makebox(0,0)[r]{0.001}}
\put(1416.0,495.0){\rule[-0.200pt]{4.818pt}{0.400pt}}
\put(220.0,591.0){\rule[-0.200pt]{4.818pt}{0.400pt}}
\put(198,591){\makebox(0,0)[r]{0.0011}}
\put(1416.0,591.0){\rule[-0.200pt]{4.818pt}{0.400pt}}
\put(220.0,686.0){\rule[-0.200pt]{4.818pt}{0.400pt}}
\put(198,686){\makebox(0,0)[r]{0.0012}}
\put(1416.0,686.0){\rule[-0.200pt]{4.818pt}{0.400pt}}
\put(220.0,782.0){\rule[-0.200pt]{4.818pt}{0.400pt}}
\put(198,782){\makebox(0,0)[r]{0.0013}}
\put(1416.0,782.0){\rule[-0.200pt]{4.818pt}{0.400pt}}
\put(220.0,877.0){\rule[-0.200pt]{4.818pt}{0.400pt}}
\put(198,877){\makebox(0,0)[r]{0.0014}}
\put(1416.0,877.0){\rule[-0.200pt]{4.818pt}{0.400pt}}
\put(220.0,113.0){\rule[-0.200pt]{0.400pt}{4.818pt}}
\put(220,68){\makebox(0,0){0}}
\put(220.0,857.0){\rule[-0.200pt]{0.400pt}{4.818pt}}
\put(355.0,113.0){\rule[-0.200pt]{0.400pt}{4.818pt}}
\put(355,68){\makebox(0,0){20}}
\put(355.0,857.0){\rule[-0.200pt]{0.400pt}{4.818pt}}
\put(490.0,113.0){\rule[-0.200pt]{0.400pt}{4.818pt}}
\put(490,68){\makebox(0,0){40}}
\put(490.0,857.0){\rule[-0.200pt]{0.400pt}{4.818pt}}
\put(625.0,113.0){\rule[-0.200pt]{0.400pt}{4.818pt}}
\put(625,68){\makebox(0,0){60}}
\put(625.0,857.0){\rule[-0.200pt]{0.400pt}{4.818pt}}
\put(760.0,113.0){\rule[-0.200pt]{0.400pt}{4.818pt}}
\put(760,68){\makebox(0,0){80}}
\put(760.0,857.0){\rule[-0.200pt]{0.400pt}{4.818pt}}
\put(896.0,113.0){\rule[-0.200pt]{0.400pt}{4.818pt}}
\put(896,68){\makebox(0,0){100}}
\put(896.0,857.0){\rule[-0.200pt]{0.400pt}{4.818pt}}
\put(1031.0,113.0){\rule[-0.200pt]{0.400pt}{4.818pt}}
\put(1031,68){\makebox(0,0){120}}
\put(1031.0,857.0){\rule[-0.200pt]{0.400pt}{4.818pt}}
\put(1166.0,113.0){\rule[-0.200pt]{0.400pt}{4.818pt}}
\put(1166,68){\makebox(0,0){140}}
\put(1166.0,857.0){\rule[-0.200pt]{0.400pt}{4.818pt}}
\put(1301.0,113.0){\rule[-0.200pt]{0.400pt}{4.818pt}}
\put(1301,68){\makebox(0,0){160}}
\put(1301.0,857.0){\rule[-0.200pt]{0.400pt}{4.818pt}}
\put(1436.0,113.0){\rule[-0.200pt]{0.400pt}{4.818pt}}
\put(1436,68){\makebox(0,0){180}}
\put(1436.0,857.0){\rule[-0.200pt]{0.400pt}{4.818pt}}
\put(220.0,113.0){\rule[-0.200pt]{292.934pt}{0.400pt}}
\put(1436.0,113.0){\rule[-0.200pt]{0.400pt}{184.048pt}}
\put(220.0,877.0){\rule[-0.200pt]{292.934pt}{0.400pt}}
\put(0,495){\makebox(0,0){$d\sigma$ (pb)}}
\put(828,18){\makebox(0,0){Opening Angle in degree}}
\put(220.0,113.0){\rule[-0.200pt]{0.400pt}{184.048pt}}
\put(466,812){\makebox(0,0)[r]{1st lepton}} 
\put(466.0,812.0){\rule[-0.200pt]{15.899pt}{0.400pt}}
\put(220.0,198.0){\rule[-0.200pt]{5.782pt}{0.400pt}}
\put(244.0,198.0){\rule[-0.200pt]{0.400pt}{2.891pt}}
\put(244.0,210.0){\rule[-0.200pt]{6.022pt}{0.400pt}}
\put(269.0,198.0){\rule[-0.200pt]{0.400pt}{2.891pt}}
\put(269.0,198.0){\rule[-0.200pt]{5.782pt}{0.400pt}}
\put(293.0,198.0){\rule[-0.200pt]{0.400pt}{2.168pt}}
\put(293.0,207.0){\rule[-0.200pt]{11.804pt}{0.400pt}}
\put(342.0,207.0){\rule[-0.200pt]{0.400pt}{2.409pt}}
\put(342.0,217.0){\rule[-0.200pt]{5.782pt}{0.400pt}}
\put(366.0,216.0){\usebox{\plotpoint}}
\put(366.0,216.0){\rule[-0.200pt]{5.782pt}{0.400pt}}
\put(390.0,216.0){\rule[-0.200pt]{0.400pt}{2.891pt}}
\put(390.0,228.0){\rule[-0.200pt]{6.022pt}{0.400pt}}
\put(415.0,228.0){\rule[-0.200pt]{0.400pt}{1.204pt}}
\put(415.0,233.0){\rule[-0.200pt]{5.782pt}{0.400pt}}
\put(439.0,233.0){\rule[-0.200pt]{0.400pt}{0.723pt}}
\put(439.0,236.0){\rule[-0.200pt]{5.782pt}{0.400pt}}
\put(463.0,236.0){\rule[-0.200pt]{0.400pt}{2.650pt}}
\put(463.0,247.0){\rule[-0.200pt]{6.022pt}{0.400pt}}
\put(488.0,243.0){\rule[-0.200pt]{0.400pt}{0.964pt}}
\put(488.0,243.0){\rule[-0.200pt]{5.782pt}{0.400pt}}
\put(512.0,243.0){\rule[-0.200pt]{0.400pt}{0.723pt}}
\put(512.0,246.0){\rule[-0.200pt]{5.782pt}{0.400pt}}
\put(536.0,246.0){\rule[-0.200pt]{0.400pt}{3.613pt}}
\put(536.0,261.0){\rule[-0.200pt]{5.782pt}{0.400pt}}
\put(560.0,261.0){\rule[-0.200pt]{0.400pt}{1.927pt}}
\put(560.0,269.0){\rule[-0.200pt]{6.022pt}{0.400pt}}
\put(585.0,269.0){\rule[-0.200pt]{0.400pt}{8.913pt}}
\put(585.0,306.0){\rule[-0.200pt]{5.782pt}{0.400pt}}
\put(609.0,305.0){\usebox{\plotpoint}}
\put(609.0,305.0){\rule[-0.200pt]{5.782pt}{0.400pt}}
\put(633.0,299.0){\rule[-0.200pt]{0.400pt}{1.445pt}}
\put(633.0,299.0){\rule[-0.200pt]{6.022pt}{0.400pt}}
\put(658.0,299.0){\rule[-0.200pt]{0.400pt}{4.095pt}}
\put(658.0,316.0){\rule[-0.200pt]{5.782pt}{0.400pt}}
\put(682.0,316.0){\rule[-0.200pt]{0.400pt}{3.854pt}}
\put(682.0,332.0){\rule[-0.200pt]{5.782pt}{0.400pt}}
\put(706.0,331.0){\usebox{\plotpoint}}
\put(706.0,331.0){\rule[-0.200pt]{6.022pt}{0.400pt}}
\put(731.0,331.0){\rule[-0.200pt]{0.400pt}{8.191pt}}
\put(731.0,365.0){\rule[-0.200pt]{5.782pt}{0.400pt}}
\put(755.0,357.0){\rule[-0.200pt]{0.400pt}{1.927pt}}
\put(755.0,357.0){\rule[-0.200pt]{5.782pt}{0.400pt}}
\put(779.0,357.0){\rule[-0.200pt]{0.400pt}{11.563pt}}
\put(779.0,405.0){\rule[-0.200pt]{6.022pt}{0.400pt}}
\put(804.0,399.0){\rule[-0.200pt]{0.400pt}{1.445pt}}
\put(804.0,399.0){\rule[-0.200pt]{5.782pt}{0.400pt}}
\put(828.0,399.0){\rule[-0.200pt]{0.400pt}{2.891pt}}
\put(828.0,411.0){\rule[-0.200pt]{5.782pt}{0.400pt}}
\put(852.0,411.0){\rule[-0.200pt]{0.400pt}{6.022pt}}
\put(852.0,436.0){\rule[-0.200pt]{6.022pt}{0.400pt}}
\put(877.0,436.0){\rule[-0.200pt]{0.400pt}{2.168pt}}
\put(877.0,445.0){\rule[-0.200pt]{5.782pt}{0.400pt}}
\put(901.0,445.0){\rule[-0.200pt]{0.400pt}{3.854pt}}
\put(901.0,461.0){\rule[-0.200pt]{5.782pt}{0.400pt}}
\put(925.0,461.0){\rule[-0.200pt]{0.400pt}{3.854pt}}
\put(925.0,477.0){\rule[-0.200pt]{6.022pt}{0.400pt}}
\put(950.0,477.0){\rule[-0.200pt]{0.400pt}{9.154pt}}
\put(950.0,515.0){\rule[-0.200pt]{5.782pt}{0.400pt}}
\put(974.0,515.0){\rule[-0.200pt]{0.400pt}{0.723pt}}
\put(974.0,518.0){\rule[-0.200pt]{5.782pt}{0.400pt}}
\put(998.0,518.0){\rule[-0.200pt]{0.400pt}{9.636pt}}
\put(998.0,558.0){\rule[-0.200pt]{6.022pt}{0.400pt}}
\put(1023.0,543.0){\rule[-0.200pt]{0.400pt}{3.613pt}}
\put(1023.0,543.0){\rule[-0.200pt]{5.782pt}{0.400pt}}
\put(1047.0,543.0){\rule[-0.200pt]{0.400pt}{11.563pt}}
\put(1047.0,591.0){\rule[-0.200pt]{5.782pt}{0.400pt}}
\put(1071.0,591.0){\rule[-0.200pt]{0.400pt}{1.686pt}}
\put(1071.0,598.0){\rule[-0.200pt]{6.022pt}{0.400pt}}
\put(1096.0,598.0){\usebox{\plotpoint}}
\put(1096.0,599.0){\rule[-0.200pt]{5.782pt}{0.400pt}}
\put(1120.0,599.0){\rule[-0.200pt]{0.400pt}{11.081pt}}
\put(1120.0,645.0){\rule[-0.200pt]{5.782pt}{0.400pt}}
\put(1144.0,645.0){\rule[-0.200pt]{0.400pt}{2.168pt}}
\put(1144.0,654.0){\rule[-0.200pt]{5.782pt}{0.400pt}}
\put(1168.0,654.0){\rule[-0.200pt]{0.400pt}{5.059pt}}
\put(1168.0,675.0){\rule[-0.200pt]{11.804pt}{0.400pt}}
\put(1217.0,675.0){\rule[-0.200pt]{0.400pt}{6.745pt}}
\put(1217.0,703.0){\rule[-0.200pt]{5.782pt}{0.400pt}}
\put(1241.0,703.0){\rule[-0.200pt]{0.400pt}{5.059pt}}
\put(1241.0,724.0){\rule[-0.200pt]{6.022pt}{0.400pt}}
\put(1266.0,724.0){\rule[-0.200pt]{0.400pt}{5.300pt}}
\put(1266.0,746.0){\rule[-0.200pt]{5.782pt}{0.400pt}}
\put(1290.0,735.0){\rule[-0.200pt]{0.400pt}{2.650pt}}
\put(1290.0,735.0){\rule[-0.200pt]{5.782pt}{0.400pt}}
\put(1314.0,735.0){\rule[-0.200pt]{0.400pt}{5.300pt}}
\put(1314.0,757.0){\rule[-0.200pt]{6.022pt}{0.400pt}}
\put(1339.0,738.0){\rule[-0.200pt]{0.400pt}{4.577pt}}
\put(1339.0,738.0){\rule[-0.200pt]{5.782pt}{0.400pt}}
\put(1363.0,738.0){\rule[-0.200pt]{0.400pt}{9.636pt}}
\put(1363.0,778.0){\rule[-0.200pt]{5.782pt}{0.400pt}}
\put(1387.0,770.0){\rule[-0.200pt]{0.400pt}{1.927pt}}
\put(1387.0,770.0){\rule[-0.200pt]{6.022pt}{0.400pt}}
\put(1412.0,770.0){\rule[-0.200pt]{0.400pt}{6.504pt}}
\put(1412.0,797.0){\rule[-0.200pt]{5.782pt}{0.400pt}}
\sbox{\plotpoint}{\rule[-0.400pt]{0.800pt}{0.800pt}}%
\put(466,767){\makebox(0,0)[r]{2nd lepton}}
\put(466.0,767.0){\rule[-0.400pt]{15.899pt}{0.800pt}}
\put(220.0,228.0){\rule[-0.400pt]{5.782pt}{0.800pt}}
\put(244.0,218.0){\rule[-0.400pt]{0.800pt}{2.409pt}}
\put(244.0,218.0){\rule[-0.400pt]{6.022pt}{0.800pt}}
\put(269.0,218.0){\rule[-0.400pt]{0.800pt}{2.891pt}}
\put(269.0,230.0){\rule[-0.400pt]{5.782pt}{0.800pt}}
\put(293.0,230.0){\usebox{\plotpoint}}
\put(293.0,232.0){\rule[-0.400pt]{5.782pt}{0.800pt}}
\put(317.0,232.0){\rule[-0.400pt]{0.800pt}{0.964pt}}
\put(317.0,236.0){\rule[-0.400pt]{6.022pt}{0.800pt}}
\put(342.0,236.0){\rule[-0.400pt]{0.800pt}{1.927pt}}
\put(342.0,244.0){\rule[-0.400pt]{5.782pt}{0.800pt}}
\put(366.0,244.0){\usebox{\plotpoint}}
\put(366.0,246.0){\rule[-0.400pt]{5.782pt}{0.800pt}}
\put(390.0,241.0){\rule[-0.400pt]{0.800pt}{1.204pt}}
\put(390.0,241.0){\rule[-0.400pt]{6.022pt}{0.800pt}}
\put(415.0,241.0){\rule[-0.400pt]{0.800pt}{3.132pt}}
\put(415.0,254.0){\rule[-0.400pt]{5.782pt}{0.800pt}}
\put(439.0,251.0){\usebox{\plotpoint}}
\put(439.0,251.0){\rule[-0.400pt]{5.782pt}{0.800pt}}
\put(463.0,251.0){\rule[-0.400pt]{0.800pt}{4.818pt}}
\put(463.0,271.0){\rule[-0.400pt]{6.022pt}{0.800pt}}
\put(488.0,271.0){\usebox{\plotpoint}}
\put(488.0,272.0){\rule[-0.400pt]{5.782pt}{0.800pt}}
\put(512.0,269.0){\usebox{\plotpoint}}
\put(512.0,269.0){\rule[-0.400pt]{5.782pt}{0.800pt}}
\put(536.0,269.0){\rule[-0.400pt]{0.800pt}{4.818pt}}
\put(536.0,289.0){\rule[-0.400pt]{5.782pt}{0.800pt}}
\put(560.0,289.0){\rule[-0.400pt]{0.800pt}{3.373pt}}
\put(560.0,303.0){\rule[-0.400pt]{6.022pt}{0.800pt}}
\put(585.0,303.0){\usebox{\plotpoint}}
\put(585.0,305.0){\rule[-0.400pt]{5.782pt}{0.800pt}}
\put(609.0,305.0){\rule[-0.400pt]{0.800pt}{4.095pt}}
\put(609.0,322.0){\rule[-0.400pt]{5.782pt}{0.800pt}}
\put(633.0,322.0){\rule[-0.400pt]{0.800pt}{3.132pt}}
\put(633.0,335.0){\rule[-0.400pt]{6.022pt}{0.800pt}}
\put(658.0,335.0){\rule[-0.400pt]{0.800pt}{2.168pt}}
\put(658.0,344.0){\rule[-0.400pt]{5.782pt}{0.800pt}}
\put(682.0,344.0){\rule[-0.400pt]{0.800pt}{4.095pt}}
\put(682.0,361.0){\rule[-0.400pt]{5.782pt}{0.800pt}}
\put(706.0,361.0){\usebox{\plotpoint}}
\put(706.0,363.0){\rule[-0.400pt]{6.022pt}{0.800pt}}
\put(731.0,363.0){\rule[-0.400pt]{0.800pt}{2.891pt}}
\put(731.0,375.0){\rule[-0.400pt]{5.782pt}{0.800pt}}
\put(755.0,375.0){\rule[-0.400pt]{0.800pt}{6.504pt}}
\put(755.0,402.0){\rule[-0.400pt]{5.782pt}{0.800pt}}
\put(779.0,402.0){\rule[-0.400pt]{0.800pt}{6.263pt}}
\put(779.0,428.0){\rule[-0.400pt]{6.022pt}{0.800pt}}
\put(804.0,409.0){\rule[-0.400pt]{0.800pt}{4.577pt}}
\put(804.0,409.0){\rule[-0.400pt]{5.782pt}{0.800pt}}
\put(828.0,409.0){\rule[-0.400pt]{0.800pt}{6.504pt}}
\put(828.0,436.0){\rule[-0.400pt]{5.782pt}{0.800pt}}
\put(852.0,436.0){\rule[-0.400pt]{0.800pt}{12.045pt}}
\put(852.0,486.0){\rule[-0.400pt]{6.022pt}{0.800pt}}
\put(877.0,486.0){\rule[-0.400pt]{0.800pt}{1.204pt}}
\put(877.0,491.0){\rule[-0.400pt]{5.782pt}{0.800pt}}
\put(901.0,489.0){\usebox{\plotpoint}}
\put(901.0,489.0){\rule[-0.400pt]{5.782pt}{0.800pt}}
\put(925.0,489.0){\rule[-0.400pt]{0.800pt}{14.213pt}}
\put(925.0,548.0){\rule[-0.400pt]{6.022pt}{0.800pt}}
\put(950.0,538.0){\rule[-0.400pt]{0.800pt}{2.409pt}}
\put(950.0,538.0){\rule[-0.400pt]{5.782pt}{0.800pt}}
\put(974.0,538.0){\rule[-0.400pt]{0.800pt}{3.854pt}}
\put(974.0,554.0){\rule[-0.400pt]{5.782pt}{0.800pt}}
\put(998.0,554.0){\rule[-0.400pt]{0.800pt}{6.504pt}}
\put(998.0,581.0){\rule[-0.400pt]{6.022pt}{0.800pt}}
\put(1023.0,581.0){\rule[-0.400pt]{0.800pt}{1.686pt}}
\put(1023.0,588.0){\rule[-0.400pt]{5.782pt}{0.800pt}}
\put(1047.0,588.0){\rule[-0.400pt]{0.800pt}{7.709pt}}
\put(1047.0,620.0){\rule[-0.400pt]{5.782pt}{0.800pt}}
\put(1071.0,620.0){\rule[-0.400pt]{0.800pt}{3.132pt}}
\put(1071.0,633.0){\rule[-0.400pt]{6.022pt}{0.800pt}}
\put(1096.0,633.0){\usebox{\plotpoint}}
\put(1096.0,635.0){\rule[-0.400pt]{5.782pt}{0.800pt}}
\put(1120.0,635.0){\rule[-0.400pt]{0.800pt}{6.745pt}}
\put(1120.0,663.0){\rule[-0.400pt]{5.782pt}{0.800pt}}
\put(1144.0,646.0){\rule[-0.400pt]{0.800pt}{4.095pt}}
\put(1144.0,646.0){\rule[-0.400pt]{5.782pt}{0.800pt}}
\put(1168.0,646.0){\rule[-0.400pt]{0.800pt}{4.095pt}}
\put(1168.0,663.0){\rule[-0.400pt]{6.022pt}{0.800pt}}
\put(1193.0,663.0){\rule[-0.400pt]{0.800pt}{2.409pt}}
\put(1193.0,673.0){\rule[-0.400pt]{5.782pt}{0.800pt}}
\put(1217.0,673.0){\rule[-0.400pt]{0.800pt}{5.782pt}}
\put(1217.0,697.0){\rule[-0.400pt]{5.782pt}{0.800pt}}
\put(1241.0,675.0){\rule[-0.400pt]{0.800pt}{5.300pt}}
\put(1241.0,675.0){\rule[-0.400pt]{6.022pt}{0.800pt}}
\put(1266.0,654.0){\rule[-0.400pt]{0.800pt}{5.059pt}}
\put(1266.0,654.0){\rule[-0.400pt]{5.782pt}{0.800pt}}
\put(1290.0,640.0){\rule[-0.400pt]{0.800pt}{3.373pt}}
\put(1290.0,640.0){\rule[-0.400pt]{5.782pt}{0.800pt}}
\put(1314.0,640.0){\rule[-0.400pt]{0.800pt}{0.964pt}}
\put(1314.0,644.0){\rule[-0.400pt]{6.022pt}{0.800pt}}
\put(1339.0,606.0){\rule[-0.400pt]{0.800pt}{9.154pt}}
\put(1339.0,606.0){\rule[-0.400pt]{5.782pt}{0.800pt}}
\put(1363.0,591.0){\rule[-0.400pt]{0.800pt}{3.613pt}}
\put(1363.0,591.0){\rule[-0.400pt]{5.782pt}{0.800pt}}
\put(1387.0,591.0){\rule[-0.400pt]{0.800pt}{1.686pt}}
\put(1387.0,598.0){\rule[-0.400pt]{6.022pt}{0.800pt}}
\put(1412.0,598.0){\usebox{\plotpoint}}
\put(1412.0,600.0){\rule[-0.400pt]{5.782pt}{0.800pt}}
\sbox{\plotpoint}{\rule[-0.600pt]{1.200pt}{1.200pt}}%
\put(466,722){\makebox(0,0)[r]{3rd lepton}}
\put(466.0,722.0){\rule[-0.600pt]{15.899pt}{1.200pt}}
\put(220.0,451.0){\rule[-0.600pt]{5.782pt}{1.200pt}}
\put(244.0,431.0){\rule[-0.600pt]{1.200pt}{4.818pt}}
\put(244.0,431.0){\rule[-0.600pt]{6.022pt}{1.200pt}}
\put(269.0,431.0){\rule[-0.600pt]{1.200pt}{4.095pt}}
\put(269.0,448.0){\rule[-0.600pt]{5.782pt}{1.200pt}}
\put(293.0,448.0){\rule[-0.600pt]{1.200pt}{5.300pt}}
\put(293.0,470.0){\rule[-0.600pt]{5.782pt}{1.200pt}}
\put(317.0,451.0){\rule[-0.600pt]{1.200pt}{4.577pt}}
\put(317.0,451.0){\rule[-0.600pt]{6.022pt}{1.200pt}}
\put(342.0,446.0){\rule[-0.600pt]{1.200pt}{1.204pt}}
\put(342.0,446.0){\rule[-0.600pt]{5.782pt}{1.200pt}}
\put(366.0,445.0){\usebox{\plotpoint}}
\put(366.0,445.0){\rule[-0.600pt]{5.782pt}{1.200pt}}
\put(390.0,445.0){\usebox{\plotpoint}}
\put(390.0,448.0){\rule[-0.600pt]{6.022pt}{1.200pt}}
\put(415.0,448.0){\rule[-0.600pt]{1.200pt}{6.504pt}}
\put(415.0,475.0){\rule[-0.600pt]{5.782pt}{1.200pt}}
\put(439.0,437.0){\rule[-0.600pt]{1.200pt}{9.154pt}}
\put(439.0,437.0){\rule[-0.600pt]{5.782pt}{1.200pt}}
\put(463.0,437.0){\rule[-0.600pt]{1.200pt}{3.854pt}}
\put(463.0,453.0){\rule[-0.600pt]{6.022pt}{1.200pt}}
\put(488.0,448.0){\rule[-0.600pt]{1.200pt}{1.204pt}}
\put(488.0,448.0){\rule[-0.600pt]{5.782pt}{1.200pt}}
\put(512.0,448.0){\rule[-0.600pt]{1.200pt}{7.227pt}}
\put(512.0,478.0){\rule[-0.600pt]{5.782pt}{1.200pt}}
\put(536.0,459.0){\rule[-0.600pt]{1.200pt}{4.577pt}}
\put(536.0,459.0){\rule[-0.600pt]{5.782pt}{1.200pt}}
\put(560.0,459.0){\rule[-0.600pt]{1.200pt}{4.095pt}}
\put(560.0,476.0){\rule[-0.600pt]{11.804pt}{1.200pt}}
\put(609.0,475.0){\usebox{\plotpoint}}
\put(609.0,475.0){\rule[-0.600pt]{5.782pt}{1.200pt}}
\put(633.0,475.0){\rule[-0.600pt]{1.200pt}{2.891pt}}
\put(633.0,487.0){\rule[-0.600pt]{6.022pt}{1.200pt}}
\put(658.0,479.0){\rule[-0.600pt]{1.200pt}{1.927pt}}
\put(658.0,479.0){\rule[-0.600pt]{5.782pt}{1.200pt}}
\put(682.0,473.0){\rule[-0.600pt]{1.200pt}{1.445pt}}
\put(682.0,473.0){\rule[-0.600pt]{5.782pt}{1.200pt}}
\put(706.0,473.0){\rule[-0.600pt]{1.200pt}{2.409pt}}
\put(706.0,483.0){\rule[-0.600pt]{11.804pt}{1.200pt}}
\put(755.0,472.0){\rule[-0.600pt]{1.200pt}{2.650pt}}
\put(755.0,472.0){\rule[-0.600pt]{5.782pt}{1.200pt}}
\put(779.0,472.0){\usebox{\plotpoint}}
\put(779.0,474.0){\rule[-0.600pt]{6.022pt}{1.200pt}}
\put(804.0,461.0){\rule[-0.600pt]{1.200pt}{3.132pt}}
\put(804.0,461.0){\rule[-0.600pt]{5.782pt}{1.200pt}}
\put(828.0,461.0){\rule[-0.600pt]{1.200pt}{1.204pt}}
\put(828.0,466.0){\rule[-0.600pt]{5.782pt}{1.200pt}}
\put(852.0,466.0){\rule[-0.600pt]{1.200pt}{4.577pt}}
\put(852.0,485.0){\rule[-0.600pt]{6.022pt}{1.200pt}}
\put(877.0,462.0){\rule[-0.600pt]{1.200pt}{5.541pt}}
\put(877.0,462.0){\rule[-0.600pt]{5.782pt}{1.200pt}}
\put(901.0,451.0){\rule[-0.600pt]{1.200pt}{2.650pt}}
\put(901.0,451.0){\rule[-0.600pt]{5.782pt}{1.200pt}}
\put(925.0,451.0){\usebox{\plotpoint}}
\put(925.0,452.0){\rule[-0.600pt]{6.022pt}{1.200pt}}
\put(950.0,452.0){\rule[-0.600pt]{1.200pt}{1.204pt}}
\put(950.0,457.0){\rule[-0.600pt]{5.782pt}{1.200pt}}
\put(974.0,451.0){\rule[-0.600pt]{1.200pt}{1.445pt}}
\put(974.0,451.0){\rule[-0.600pt]{5.782pt}{1.200pt}}
\put(998.0,451.0){\rule[-0.600pt]{1.200pt}{3.854pt}}
\put(998.0,467.0){\rule[-0.600pt]{6.022pt}{1.200pt}}
\put(1023.0,445.0){\rule[-0.600pt]{1.200pt}{5.300pt}}
\put(1023.0,445.0){\rule[-0.600pt]{5.782pt}{1.200pt}}
\put(1047.0,428.0){\rule[-0.600pt]{1.200pt}{4.095pt}}
\put(1047.0,428.0){\rule[-0.600pt]{5.782pt}{1.200pt}}
\put(1071.0,428.0){\usebox{\plotpoint}}
\put(1071.0,429.0){\rule[-0.600pt]{6.022pt}{1.200pt}}
\put(1096.0,417.0){\rule[-0.600pt]{1.200pt}{2.891pt}}
\put(1096.0,417.0){\rule[-0.600pt]{5.782pt}{1.200pt}}
\put(1120.0,417.0){\rule[-0.600pt]{1.200pt}{3.854pt}}
\put(1120.0,433.0){\rule[-0.600pt]{5.782pt}{1.200pt}}
\put(1144.0,415.0){\rule[-0.600pt]{1.200pt}{4.336pt}}
\put(1144.0,415.0){\rule[-0.600pt]{5.782pt}{1.200pt}}
\put(1168.0,415.0){\rule[-0.600pt]{1.200pt}{4.577pt}}
\put(1168.0,434.0){\rule[-0.600pt]{6.022pt}{1.200pt}}
\put(1193.0,420.0){\rule[-0.600pt]{1.200pt}{3.373pt}}
\put(1193.0,420.0){\rule[-0.600pt]{5.782pt}{1.200pt}}
\put(1217.0,400.0){\rule[-0.600pt]{1.200pt}{4.818pt}}
\put(1217.0,400.0){\rule[-0.600pt]{5.782pt}{1.200pt}}
\put(1241.0,400.0){\rule[-0.600pt]{1.200pt}{1.686pt}}
\put(1241.0,407.0){\rule[-0.600pt]{6.022pt}{1.200pt}}
\put(1266.0,407.0){\usebox{\plotpoint}}
\put(1266.0,409.0){\rule[-0.600pt]{5.782pt}{1.200pt}}
\put(1290.0,409.0){\rule[-0.600pt]{1.200pt}{5.059pt}}
\put(1290.0,430.0){\rule[-0.600pt]{5.782pt}{1.200pt}}
\put(1314.0,404.0){\rule[-0.600pt]{1.200pt}{6.263pt}}
\put(1314.0,404.0){\rule[-0.600pt]{6.022pt}{1.200pt}}
\put(1339.0,402.0){\usebox{\plotpoint}}
\put(1339.0,402.0){\rule[-0.600pt]{5.782pt}{1.200pt}}
\put(1363.0,402.0){\rule[-0.600pt]{1.200pt}{2.409pt}}
\put(1363.0,412.0){\rule[-0.600pt]{5.782pt}{1.200pt}}
\put(1387.0,411.0){\usebox{\plotpoint}}
\put(1387.0,411.0){\rule[-0.600pt]{6.022pt}{1.200pt}}
\put(1412.0,408.0){\usebox{\plotpoint}}
\put(1412.0,408.0){\rule[-0.600pt]{5.782pt}{1.200pt}}
\end{picture}
\vskip 1.0 in
\begin{center}
\bf{Figure 4}
\end{center}
\vskip .1in
\noindent
Distribution of azimuthal opening angle  of three leptons for the 
process $e^-\gamma\longrightarrow e^-e^+e^-$ + $E{\!\!\!\!/_T}$~~, for 
$M$ = 100 TeV, $\mu$ = 300 , tan$\beta$ = 2. All cuts except the 
azimuthal opening angle cut are as discussed in the text; the 
azimuthal opening angle cut has been put to zero. The lepton 
number assignment is the same as in Figures 3(a) and 3(b). 
\end{document}